\documentclass[11pt,a4paper]{article}
\usepackage{jcappub,epsfig,multicol,bbm,latexsym,graphicx,subfigure}

\def\ga   {\gamma}
\def\Ga   {\Gamma}

\def\La   {\Lambda}

\def\lee { \left( }
\def\rii { \right) }

\def\xp {\chi^\prime}
\def\mx {m_{\chi}}
\def\mxp {m_{\chi^{\prime}}}

\def\tev {\,\mathrm{TeV}}

\begin{document}

{\small
\begin{flushright}
IPMU16-0124, DO-TH 16/23\\
\end{flushright} }

\title{AMS-02 Positron Excess and Indirect Detection of Three-body Decaying Dark Matter}
\author[a]{Hsin-Chia Cheng}
\author[b]{Wei-Chih Huang}
\author[c]{Xiaoyuan Huang}
\author[d,e]{Ian Low}
\author[f]{Yue-Lin Sming Tsai}
\author[g,h]{Qiang Yuan}

\affiliation[a]{Department of Physics, University of California, Davis, CA 95616, USA}
\affiliation[b]{Fakult\"at f\"ur Physik, Technische Universit\"at Dortmund,
44221 Dortmund, Germany}
\affiliation[c]{Physik-Department T30d, Technische Universit\"at M\"unchen, James-Franck-Stra\ss{}e, D-85748 Garching, Germany}
\affiliation[d]{High Energy Physics Division, Argonne National Laboratory,
Argonne, IL 60439, USA}
\affiliation[e]{Department of Physics and Astronomy, Northwestern University, Evanston, IL 60208, USA}
\affiliation[f]{Kavli IPMU (WPI), University of Tokyo, Kashiwa, Chiba 277-8583, Japan}
\affiliation[g]{Key Laboratory of Dark Matter and Space Astronomy, Purple
Mountain Observatory, Chinese Academy of Sciences, Nanjing 210008, China}
\affiliation[h]{School of Astronomy and Space Science, University of Science and Technology of China, Hefei, Anhui 230026, China}
\emailAdd{cheng@physics.ucdavis.edu}
\emailAdd{wei-chih.huang@tu-dortmund.de}
\emailAdd{huangxiaoyuan@gmail.com}
\emailAdd{ilow@northwestern.edu}
\emailAdd{smingtsai@gmail.com}
\emailAdd{yuanq@pmo.ac.cn}

%\date{\today}

\abstract{We consider indirect detection of meta-stable dark matter 
particles decaying into a stable neutral particle and a pair of standard 
model fermions. Due to the softer energy spectra from the three-body decay, 
such models could potentially explain the AMS-02 positron excess without 
being constrained by the Fermi-LAT gamma-ray data and the cosmic ray 
anti-proton measurements. We scrutinize over different final state 
fermions, paying special attention to handling of the cosmic ray 
background and including various contributions from cosmic ray propagation 
with the help of the \textsc{LikeDM} package. It is found that primary 
decays into an electron-positron pair and a stable neutral particle 
could give rise to the AMS-02 positron excess and, at the same time, 
stay unscathed against the gamma-ray and anti-proton constraints. 
Decays to a muon pair or a mixed flavor electron-muon pair may also
be viable depending on the propagation models. Decays to all other 
standard model fermions are severely disfavored.}


\arxivnumber{1608.06382}

\maketitle

%%#######################################################%%
\section{Introduction \label{section:1}}
%%#######################################################%%

Although the gravitational evidence of the existence of Dark Matter~(DM) 
is cogent, the direct detection of DM particles is still far from conclusive. 
Nevertheless, several indirect detection experiments reveal intriguing excesses 
in the cosmic ray (CR) positron ratio~$e^+/(e^+ + e^-)$~\cite{Barwick:1997ig,
Aguilar:2007yf,Adriani:2008zr,FermiLAT:2011ab,Aguilar:2013qda,Accardo:2014lma} 
and electron (or $e^++e^-$) spectra~\cite{Chang:2008aa,Abdo:2009zk,
Aguilar:2014mma,Aguilar:2014fea}, which can not be easily accounted for 
with conventional astrophysical backgrounds. The excess could result from 
additional astrophysical sources, for example supernova remnants, pulsars 
or primary cosmic rays interacting with the interstellar medium
\cite{Hooper:2008kg,Delahaye:2008ua,Fujita:2009wk,Fan:2010yq,Serpico:2011wg,DiMauro:2014iia,Kohri:2015mga}. Alternatively, the DM annihilation or decay 
may also explain this result (see e.g., the reviews
\cite{Cirelli:2012tf,Bi:2014hpa,Gaskins:2016cha}). 

Models of annihilating DM usually requires large enhancement on the  
annihilation cross section~\cite{Bergstrom:2008gr,Barger:2008su,
Cholis:2008hb,Cirelli:2008pk,Yuan:2013eja,Jin:2013nta,Dev:2013hka,
Jin:2014ica}, which can be realized via, for example, the Sommerfeld 
enhancement~\cite{Hisano:2003ec,Hisano:2004ds,ArkaniHamed:2008qn,
Cirelli:2009uv} or the Breit-Wigner enhancement~\cite{Feldman:2008xs,
Ibe:2008ye}. In particular, these models also generate $\gamma$-rays from, 
for instance, the prompt radiation and the secondary inverse Compton 
(IC) scattering of $e^\pm$ with the photon background, as well as 
anti-protons from the final state hadronization. The strong constraints 
from PAMELA anti-proton and Fermi-LAT $\gamma$-ray have disfavored the DM 
annihilation as the explanation of the excesses~\cite{Donato:2008jk,
Papucci:2009gd,Cirelli:2009dv,Yuan:2013eja,Kopp:2013eka,Tavakoli:2013zva,
Feng:2013zca}.

In contrast, decaying DM does not require a large boost factor as long as 
the decay lifetime is around $10^{26}$ to $10^{27}$ seconds~\cite{Nardi:2008ix,
Yin:2008bs,Essig:2009jx,Meade:2009iu,Kajiyama:2013dba,Feng:2013vva,Ibe:2013jya,
Dienes:2013xff,Geng:2013nda,Ko:2014lsa}, depending on the DM mass and decay 
channels. The $\gamma$-ray bound, nevertheless, still applies in this case and 
has ruled out many decaying DM models based on two-body decays~\cite{Cirelli:2012ut,Carquin:2015uma,Regis:2015zka,Cuoco:2015rfa,Hamaguchi:2015wga}. 
On the other hand, there exists a class of decaying DM models which 
features three-body final states and, therefore, has softer decay spectra 
compared to that of the two-body decay. This class of models can be further 
categorized, based on final states, into two cases: all final states being 
standard model (SM) particles (for example, Ref.~\cite{Ibarra:2008jk,
Ibarra:2009dr,Kohri:2009yn,Carone:2011ur,Ibe:2013jya}) or one of the decay 
products is an absolutely stable neutral particle and thus 
invisible~\cite{Pospelov:2008rn,Demir:2009kc,Cheng:2010mw}. The constraints 
on the former case have been investigated in Ref.~\cite{Ibarra:2009dr,
Ibe:2013jya,Ando:2015qda,Ando:2016ang} and the constraints on the latter 
case were studied in Ref.~\cite{Cheng:2012uk}. In particular, the conclusion 
of Ref.~\cite{Cheng:2012uk} was that, while certain decay modes are strongly 
disfavored and part of the parameter spaces are ruled out, it is still 
possible to explain the electron/positron excesses within the constraints.
%A recent study on the three-body decaying gravitino DM, which decays into SM particles only and thus belongs to the first situation, shows that the Fermi-LAT extragalactic $\ga$-ray measurement excludes the DM-origin explanation of the positron excess~\cite{Carquin:2015uma}.

Given the new measurements of the positron and electron data by 
AMS-02~\cite{Aguilar:2013qda,Accardo:2014lma,Aguilar:2014mma,Aguilar:2014fea}
and the extragalactic $\gamma$-ray background (EGB) by 
Fermi-LAT~\cite{Ackermann:2014usa}, in this work we update the analysis 
done in Ref.~\cite{Cheng:2012uk} and consider decay channels, such as the 
mixed-flavor final states, that were not considered previously. In particular, 
we improve the analysis by adopting a more general statistical approach 
in dealing with the astrophysical background of CRs. For example, parameters 
in the astrophysical background are treated as nuisance parameters in 
scanning over the DM parameter space to fit the experimental data. 
Along the way we also examine whether the decaying DM model can give rise 
to the recently reported $\gamma$-ray excess at the Galactic center 
(GC; \cite{Goodenough:2009gk,Vitale:2009hr,Hooper:2010mq,Hooper:2011ti,
Cumberbatch:2010ii,Abazajian:2012pn,Huang:2013pda,Gordon:2013vta,
Hooper:2013rwa,Abazajian:2014fta,Daylan:2014rsa,Zhou:2014lva,Calore:2014xka,
Huang:2015rlu,TheFermi-LAT:2015kwa}). It turns out that the positron
excess and GC $\gamma$-ray excess may not be explained by a single DM component 
simultaneously due to different energy scales associated with them. 
Therefore, we treat the $\gamma$-ray data as a constraint, instead of trying 
to explain it.
 
The paper is organized as follows. We briefly describe the three-body
decaying DM model in Sec. 2. Then we fit the positron CR data to derive 
the best-fit DM model parameters in Sec. 3. The constraints on the model 
from the EGB and the $\gamma$-ray emission from the GC are discussed in 
Secs. 4 and 5, respectively. Finally we conclude in Sec. 6.
 
%%#######################################################%%
\section{Three-body dark matter decays \label{section:2}}
%%#######################################################%%

A meta-stable DM particle decaying to an absolutely stable neutral particle 
plus a pair of SM particles can occur naturally in supersymmetric theories 
with  conserved $R$-parity. An example is in models with Dirac neutrino 
masses, a heavier right-handed sneutrino can have a long lifetime of 
decaying to a lighter right-handed sneutrino with a pair of SM leptons 
due to the small Yukawa couplings~\cite{Pospelov:2008rn,Demir:2009kc}. 
Another possibility is that if supersymmetry (SUSY) is spontaneously broken 
in multiple sequestered sectors, then in each SUSY breaking sector there 
exists a goldstino. Only one linear combination of them is absorbed and 
becomes the longitudinal component of the gravitino. The other goldstini 
acquire masses through supergravity effects~\cite{Cheung:2010mc}. If the 
lightest supersymmetric particle is the gravitino and the next lightest 
supersymmetric particle is a goldstino, the  goldstino decays through 
dimension-8 operators to three-body final states of one gravitino and a 
pair of SM particles. It could naturally be long-lived and cosmologically 
stable~\cite{Cheng:2010mw}. 

In the following we discuss the primary and secondary energy spectra from 
models of such three-body decays, where a long-lived DM particle decays 
into a pair of SM fermions and a stable neutral particle.

%%#######################################################%%
\subsection{Spectra of three-body dark matter decays}
%%#######################################################%% 

We denote the decaying DM particle by $\chi$, which decays to a stable neutral particle $\chi'$ and a pair of SM fermions $\bar{f}f$. To explain the excesses in the CR electron and positron data, $m_{\chi} - m_{\chi'}$ needs to be roughly $\mathcal{O}(\rm{TeV})$. We therefore expect $\mx \sim m_{\chi'} \sim \mathcal{O}$(TeV) and the masses of the SM fermions are negligible compared with the energy scale of interest. The differential decay widths of the fermion and the anti-fermion are identical. We can define the normalized differential decay width as  
\begin{equation}
\frac{d W_f}{d E_f} = \frac{1}{\Ga}  \frac{d \Ga}{d E_f},
\end{equation}
where $\Ga$ is the partial decay width of $\chi$ to $\chi'$ and $E_f$ is the energy of the SM fermion $f$.

If $f$ is electron, then its spectrum from the decay is simply 
${d W_e}/{d E_e}$. On the other hand, if DM decays $100\%$ into $\mu^{\pm}$ 
and $\chi'$, then the electron spectrum per DM decay is
\begin{equation}
\frac{dN}{dE_e} \lee E^\prime_e\rii =
\int^{E^{\text{max}}_{\mu}}_{E^\prime_e} d E_{\mu} 
\frac{d W_\mu}{d E_\mu}\lee E_\mu\rii 
\frac{dN_\mu}{dE_e} \lee E_\mu , E^\prime_e\rii ,
\label{eq:secondary}
 \end{equation}
where $\frac{dN_\mu}{dE_e} \lee E_\mu , E^\prime_e\rii$ is the electron 
spectrum from the $\mu$ decay with energy $E_\mu$, and
\begin{equation}
E^{\text{max}}_\mu = \frac{m^2_\chi - m^2_{\chi'} } {  2 \mx}.
\end{equation} 

Ref.~\cite{Cheng:2012uk} considered many different three-body DM decay models, parametrized by different higher-dimensional operators. Here we compare the primary and secondary spectra of three cases: a fermion DM with 
$\overline{\chi^\prime} \gamma^5 \chi \bar{f} \gamma^5 f$, a scalar DM with 
$\chi \, \lee \chi'\rii^* \bar{f} f$, and the goldstino decay. We use the 
tables given in Ref.~\cite{Cirelli:2010xx} to compute the energy spectrum
of electrons/positrons, anti-protons, and photons which come from the decay
or hadronization of the fermion pair.

\begin{figure}[!htb]
\centering
\includegraphics[width=0.32\textwidth]{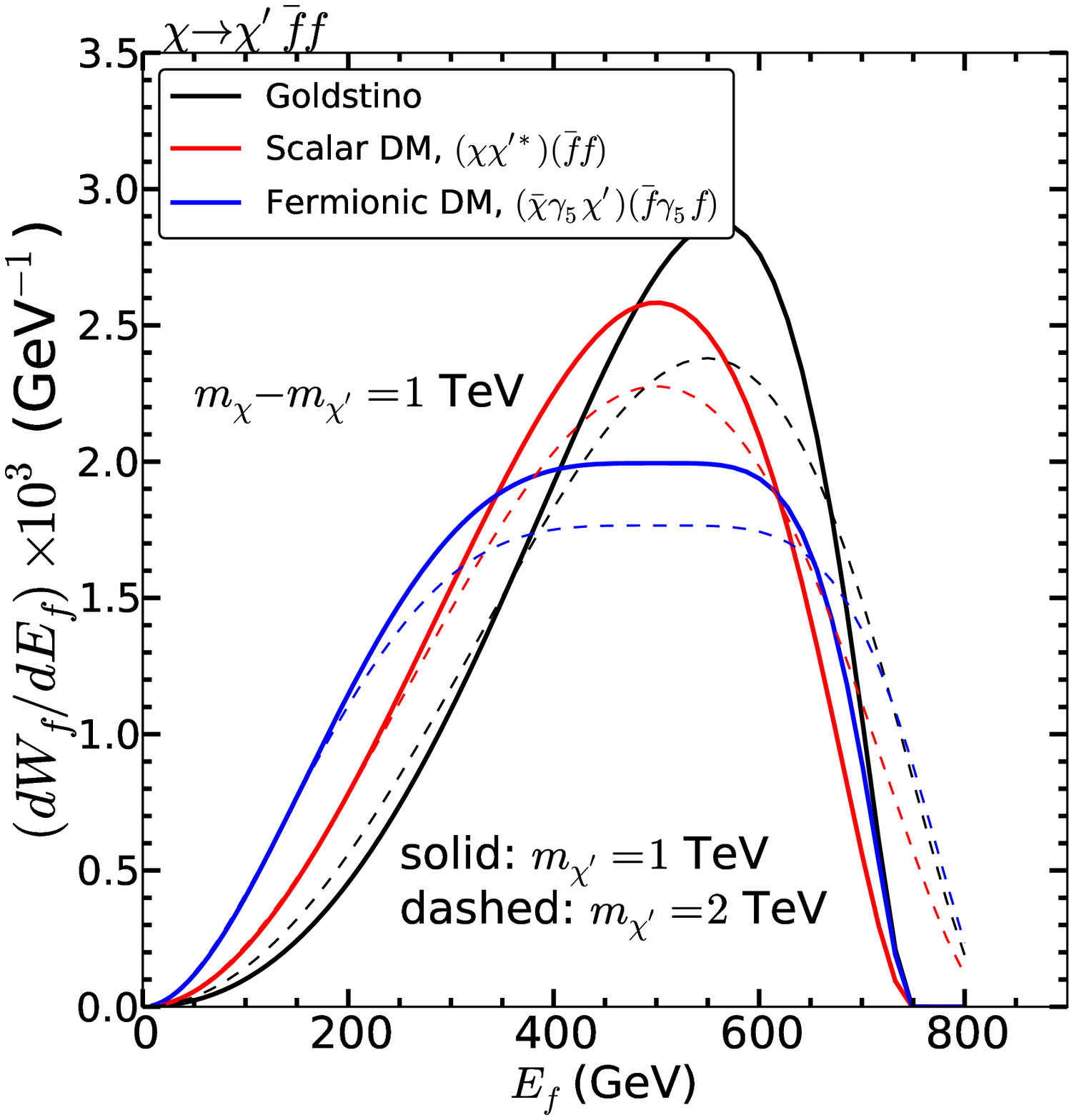}
\includegraphics[width=0.32\textwidth]{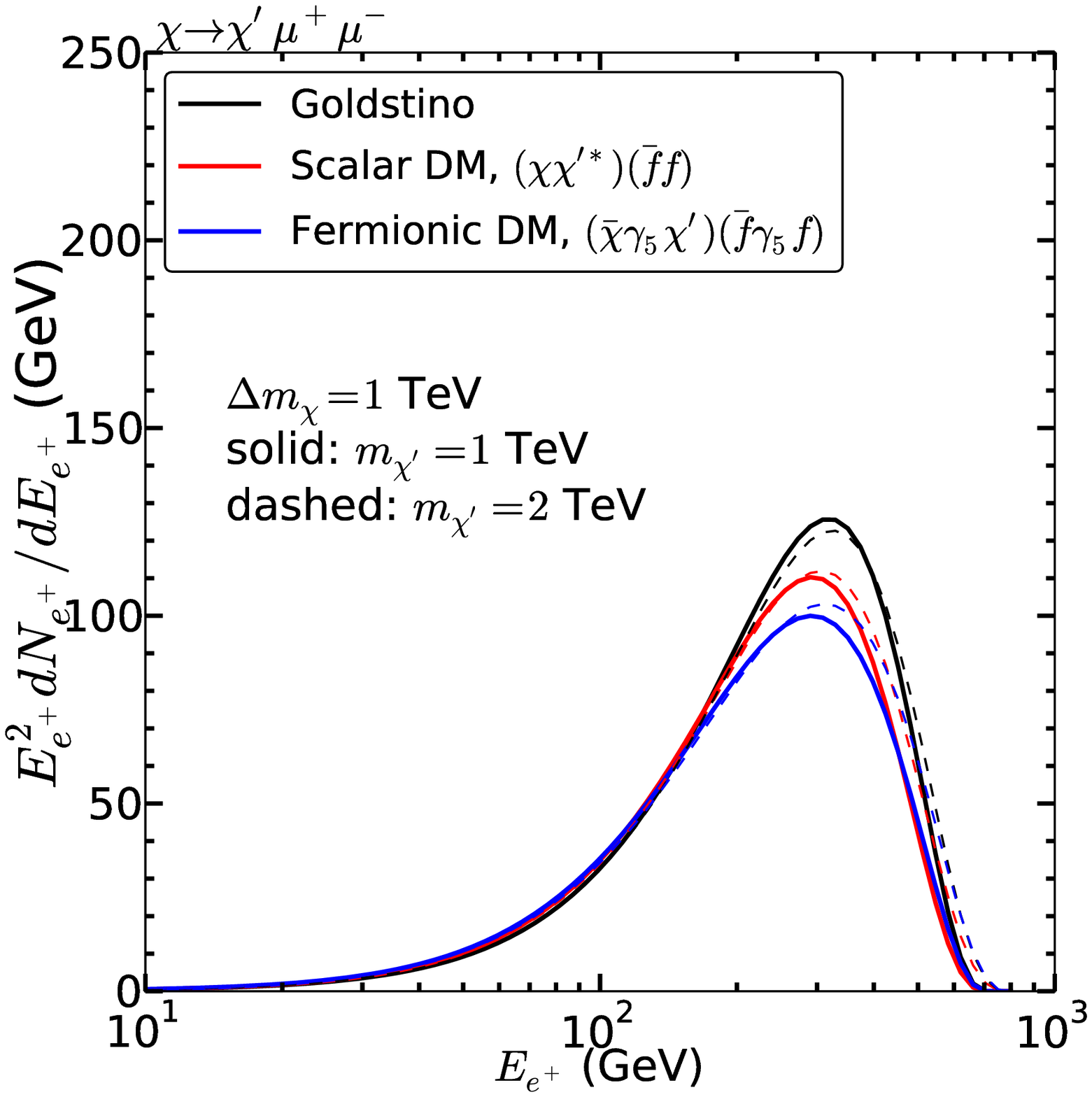}
\includegraphics[width=0.32\textwidth]{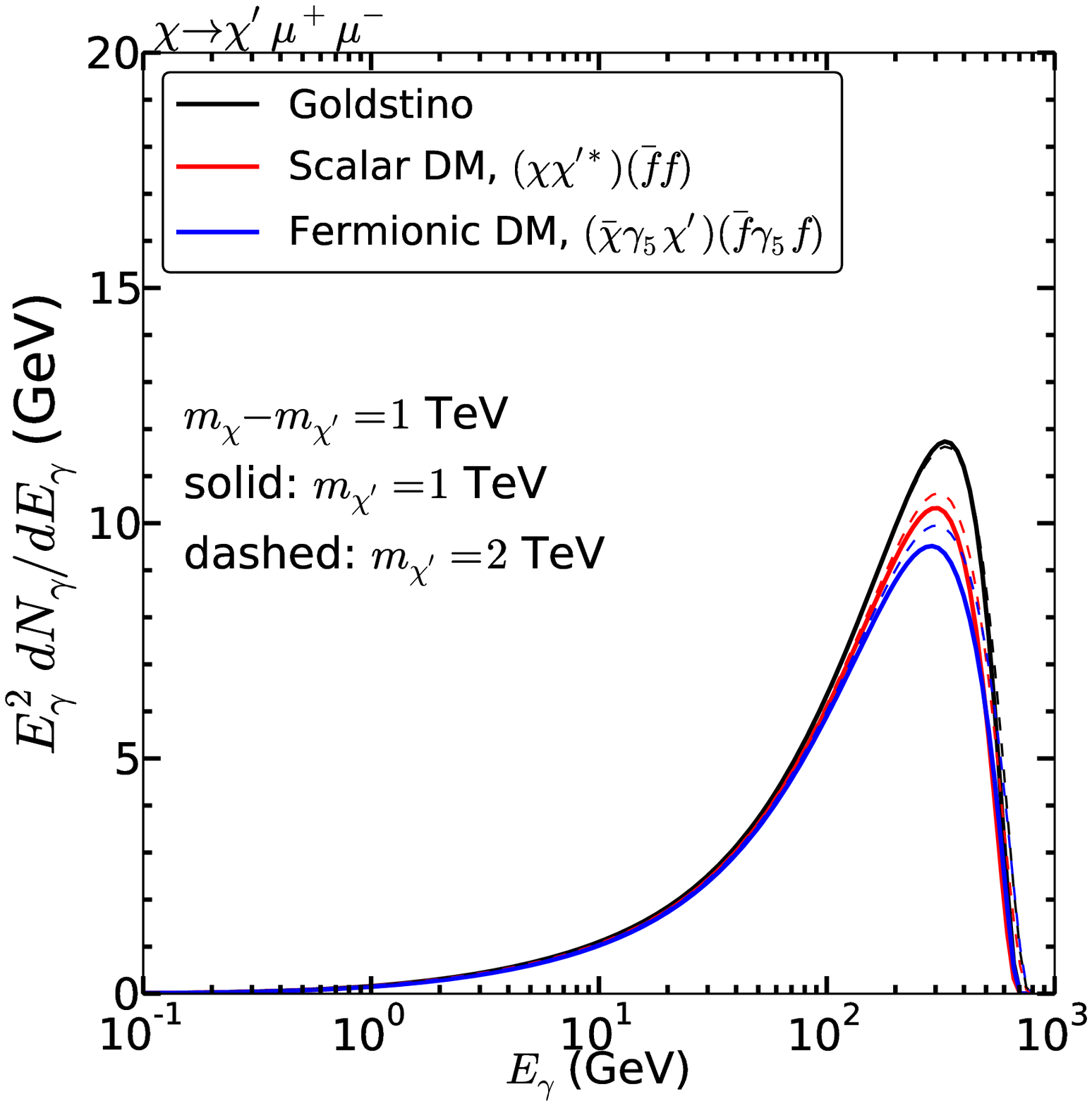}
\caption{Primary spectrum~(left), secondary electron~(central) and photon~(right) spectrum.}
\label{fig:chi_mumu}
\end{figure}

The left panel of Figure~\ref{fig:chi_mumu} shows the primary decay spectrum when the decaying DM is  
the goldstino (black), a scalar (red) and a fermion (blue). The 
dashed and solid lines correspond to $\mxp=1$ and $2$ TeV, respectively, with $\mx-\mxp$ 
fixed to be 1 TeV. The goldstino decay featuring derivative couplings has 
the hardest spectrum, while  that of the fermionic DM with axial couplings is the 
flattest. The center and right panel display the secondary electron 
and photon spectrum from final state radiation, using the decay $\chi\to \xp\mu^+\mu^-$ as the example. Although the primary spectra show some variations 
for the three types of interactions, the resulting electron and photon spectra are quite similar except for the 
peak region around 300 GeV, where both AMS-02 positron and Fermi-LAT 
$\gamma$-ray data exhibit large uncertainties.

Given the similarities of the spectra from different models, in this work we will use the intermediate energy spectrum, i.e., scalar DM 
with the scalar-type interaction $\chi \lee \chi' \rii^* \bar{f} f$, as a 
representative case for data analysis rather than studying all kinds of 
interactions.% since results will only differ slightly from operator to operator.              

%%#######################################################%%
\subsection{Scalar DM ${\chi} \lee \chi^{\prime}\rii^* \bar{f} f$}
%%#######################################################%%

The three-body decay of scalar DM can be parametrized by an effective operator
$\frac{1}{\La_\chi} {\chi} \lee \chi^{\prime}\rii^* \bar{f} f$,
where $f$ is the SM fermion. The cut-off scale ${\La_\chi}$ is treated 
as a free parameter as long as ${\La_\chi}$ is larger than $\mx$ and $\mxp$.
From the effective operator, the differential decay width into the SM 
fermion pair, $f \bar{f}$, reads
\begin{equation}
\frac{d \Gamma}{d E_f} = \frac{E^2_f}{ 32  \, \pi^3 \, m^2_\chi  \,  \La^2_\chi}
\frac{ \lee 2 \mx E_f -m^2_\chi +  m^2_{\chi'} \rii^2 } { \lee  \mx - 2 E_f \rii^2 }.
\label{eq:s_dec}
\end{equation}
The total decay width is 
\begin{equation}
\Gamma=  \frac{m^3_\chi}{768 \, \pi^3 \, \La^2_\chi } 
\left[ 1 + 9 r^2 - 9 r^4 - r^6 + 12 r^2 \lee  1 + r^2 \rii \log \lee r \rii  
\right]  ,
\label{eq:s_tdec}
\end{equation}
where $r\equiv m_{\chi'}/m_\chi$. 

The model has three parameters: $m_\chi$, $m_{\chi'}$, and $\Lambda_\chi$. 
The production rate of SM particles from DM decays mainly depends on 
$m_\chi \tau$, where $\tau =1/\Gamma$ is the lifetime of the DM particle $\chi$. 
The spectral shape of cosmic ray particles, on the other hand, is most 
sensitive to the DM mass difference $\Delta \mx=\mx-\mxp$. The dependence 
of the final cosmic ray spectra on the other combination of parameters 
is mild. Therefore we will focus on the two parameters, $m_\chi \tau$ and 
$\Delta \mx$, in our data fitting.

\section{Fitting to the charged CR data}

To infer the best-fit values of the parameters for the three-body decaying 
DM, we fit to the positron fraction \cite{Accardo:2014lma}, the total 
$e^+e^-$ spectra \cite{Aguilar:2014fea}, and the anti-proton spectrum 
\cite{Aguilar:2016kjl} measured by AMS-02. For the constraints on the
DM parameters using the preliminary AMS-02 anti-proton data, see
\cite{Giesen:2015ufa,Jin:2015sqa,Lin:2015taa}. Note that anti-protons 
can also be produced via electroweak radiative corrections even the 
decay products are leptons \cite{Ciafaloni:2010ti}. This effect has also 
been included in the fittings.

The density profile of the DM is assumed to be the Navarro-Frenk-White 
profile \cite{Navarro:1996gj}
\begin{equation}
\rho(r)=\frac{\rho_s}{(r/r_s)(1+r/r_s)^2}
\end{equation}
with $r_s=20$ kpc and $\rho_s=0.35$ GeV cm$^{-3}$ which gives a local
density of $\sim 0.4$ GeV cm$^{-3}$. The electron/positron (or anti-proton) 
source function before propagation is 
\begin{equation}
q(E,{\bf x})=\frac{1}{m_{\chi}\tau}\left(\sum_iB_i\frac{dN}{dE_i}
\right)\times\rho({\bf x}),
\end{equation}
where $\tau$ is the lifetime of the DM particle.
%\mkblue{as function of $m_{\chi}$, $m_{\chi^\prime}$, and $\La_\chi$}. 
The parameter $B_i$ is the branching 
ratio of decay channel $i$, $dN/dE_i$ is the production spectrum of 
the $e^+e^-$ (or $\bar{p}$) for channel $i$, and the summation is over 
all decay channels producing $e^+e^-$ (or $\bar{p}$). 

The propagation of electrons/positrons and anti-protons in the Milky Way
is based on the numerical tool GALPROP \cite{Strong:1998pw,Strong:2007nh}.
We adopt a fast approach using the so-called ``Green's function'' method 
which has been used in the \textsc{LikeDM} package \cite{Huang:2016pxg}.
Given the spatial distribution of the CR sources, the Green's function
of ``$\delta$-function'' kernels for a series of energies are computed 
using GALPROP. Then the final propagated spectrum of any injection
spectrum can be simply obtained via a weighted summation of the Green's
functions, where the weights are the coefficients of the kernels used to 
match the injection spectrum. We compute tables of the Green's functions
for various propagation parameters and source distributions (including
DM annihilation and decay, as well as background CR source). For more 
details we refer the readers to Ref. \cite{Huang:2016pxg}. In this work
we adopt the diffusive reacceleration propagation model. As a benchmark
parameter setting (shown in bold in Table \ref{table:prop}), we adopt 
the second group of propagation parameters given in Ref. 
\cite{Ackermann:2012rg}. To get an idea about the possible uncertainties 
of the propagation parameters, we will also consider the 1st and 6th 
groups of propagation parameters in Ref.~\cite{Ackermann:2012rg}, which 
have $z_h=2$ and 15 kpc and roughly correspond to the minimum (MIN) and 
maximum (MAX) propagation halos~\cite{Donato:2003xg,Delahaye:2007fr}. 
The detailed propagation parameters are given in Table \ref{table:prop}.

\begin{table}[!htb]
\centering
\caption{Propagation parameters}
\begin{tabular}{ccccc}
\hline \hline
No. & $D_0^*$ & $\delta^*$ & $z_h$ & $v_A$ \\
 & ($10^{28}$cm$^2$\,s$^{-1}$) & & (kpc) & (km s$^{-1}$)\\
\hline
1 & 2.7 & 0.33 & 2 & 35.0  \\
{\bf 2} & {\bf 5.3} & {\bf 0.33} & {\bf 4} & {\bf 33.5}  \\
6 & 10.0 & 0.33 & 15 & 26.3  \\
\hline
\hline
\end{tabular}\vspace{3mm}\\
$^*$The diffusion coefficient is defined as $D(R)=D_0(R/4\,{\rm GV})
^{\delta}$, with $R$ being the rigidity of particles.\\
\label{table:prop}
\end{table}

The backgrounds include primary electrons from CR acceleration sources, 
and secondary electrons/positrons and anti-protons from interactions 
between primary CR nuclei and the interstellar medium. The injection
spectra of CR nuclei and electrons are assumed to be three-segment broken 
power-law functions \cite{Yuan:2013eba,Yuan:2014pka}. The injection 
spectral parameters of primary electrons are determined simultaneously
together with the DM parameters in the global fitting. The power-law
indices and break rigidities of nuclei are determined by fitting
to the measured proton spectra \cite{Adriani:2011cu,Ahn:2010gv}, which
can be found in Ref. \cite{Yuan:2014pka}. Given the primary nuclei 
spectra, the secondary $e^+e^-$ and $\bar{p}$ fluxes can then be obtained. 
However, they can not be predicted precisely enough due to the 
following uncertainties: 1) the propagation parameters are determined 
by the Boron-to-Carbon ratios which have uncertainties, 2) there are 
uncertainties of the hadronic interaction cross sections to produce
positrons and antiprotons, and 3) there could be large fluctuations 
of the electron/positron intensities in the Milky Way due to their
shorter propagation ranges than nuclei. 
To take into account such uncertainties in predicting the secondary
$e^+e^-$ and $\bar{p}$ fluxes, we assume two free re-scaling constants 
$\alpha_{e}$ and $\alpha_{pb}$ in the fitting. Note that in
principle some of those uncertainties could be energy-dependent, and
hence the renormalization factors may not simply be constant. For 
simplicity, we will not pursue this complication here. 

The low energy charged particles are modulated by the solar wind and
its associated magnetic field after they enter the heliosphere. This solar 
modulation effect depends on the solar activities and varies in the solar 
cycle. In this work we treat the modulation effect by the force-field 
approximation \cite{Gleeson:1968zza}, with two modulation potentials 
$\Phi_{e}$ and $\Phi_{pb}$ for $e^+e^-$ and $\bar{p}$, respectively. 

The DM contribution depends on three model parameters. However, as 
we argued in Sec.~\ref{section:2}, the total flux mostly depends on 
$m_{\chi}\tau$ and the spectra of the electrons and positrons depend 
mostly on $\Delta m_{\chi}$. To reduce the number of free parameters and 
to extract the most relevant information, we simply fix 
$m_{\chi^\prime}=1\tev$ in our fitting. In total, we have 12 parameters 
in fitting the spectra: 6 for primary electrons (three power-law indices, 
two break rigidities, and one normalization), 2 for normalizations of 
secondary $e^+e^-$ and $\bar{p}$, 2 for solar modulation, and 2 for the 
DM ($m_{\chi}\tau$, $\Delta m_{\chi}$). We use the Markov Chain Monte 
Carlo (MCMC) method for the fitting \cite{Lewis:2002ah,Liu:2011re}.

\begin{figure}[!htb]
\centering
\includegraphics[width=0.32\textwidth]{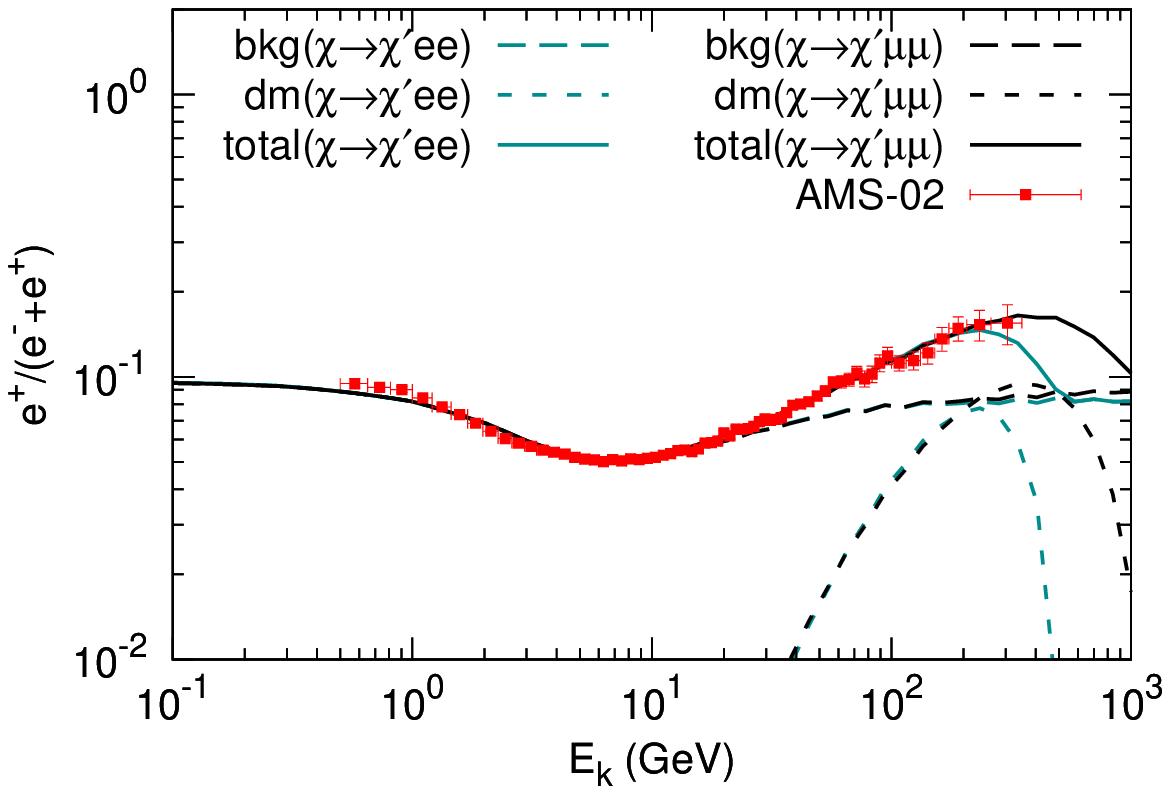}
\includegraphics[width=0.32\textwidth]{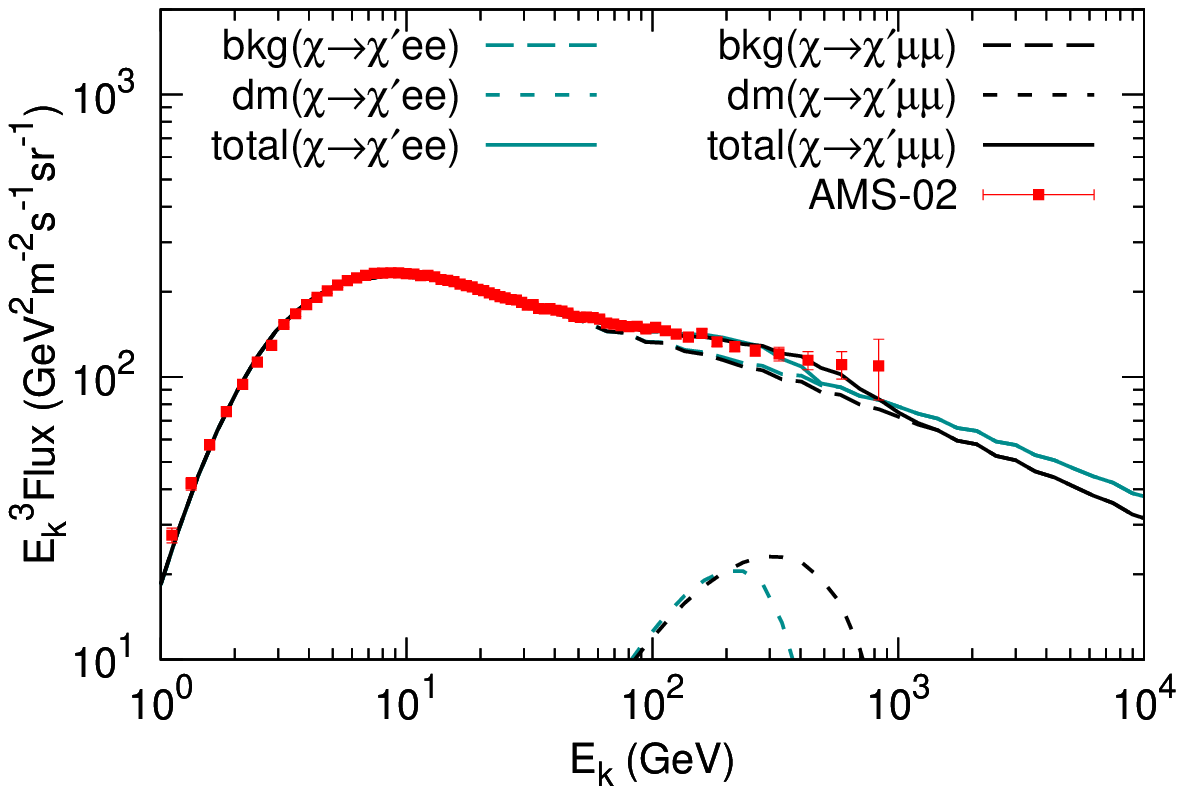}
\includegraphics[width=0.32\textwidth]{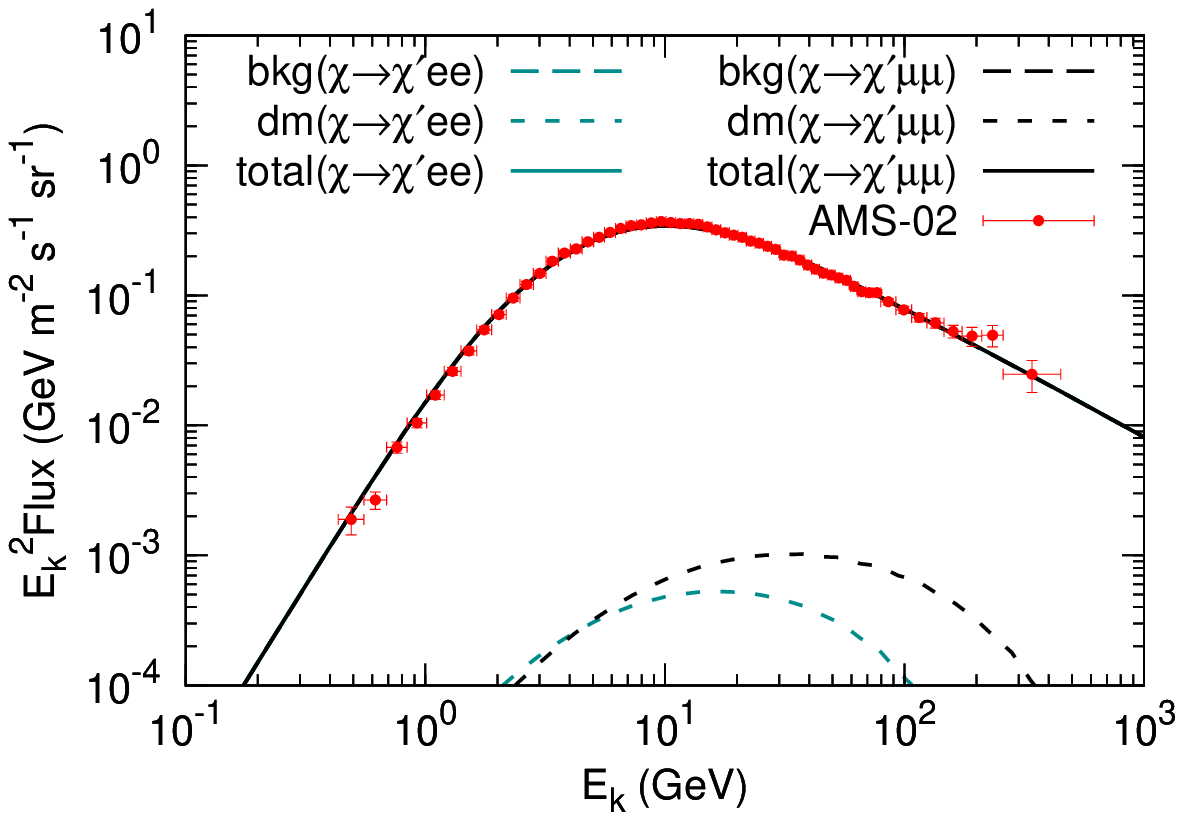}
\includegraphics[width=0.32\textwidth]{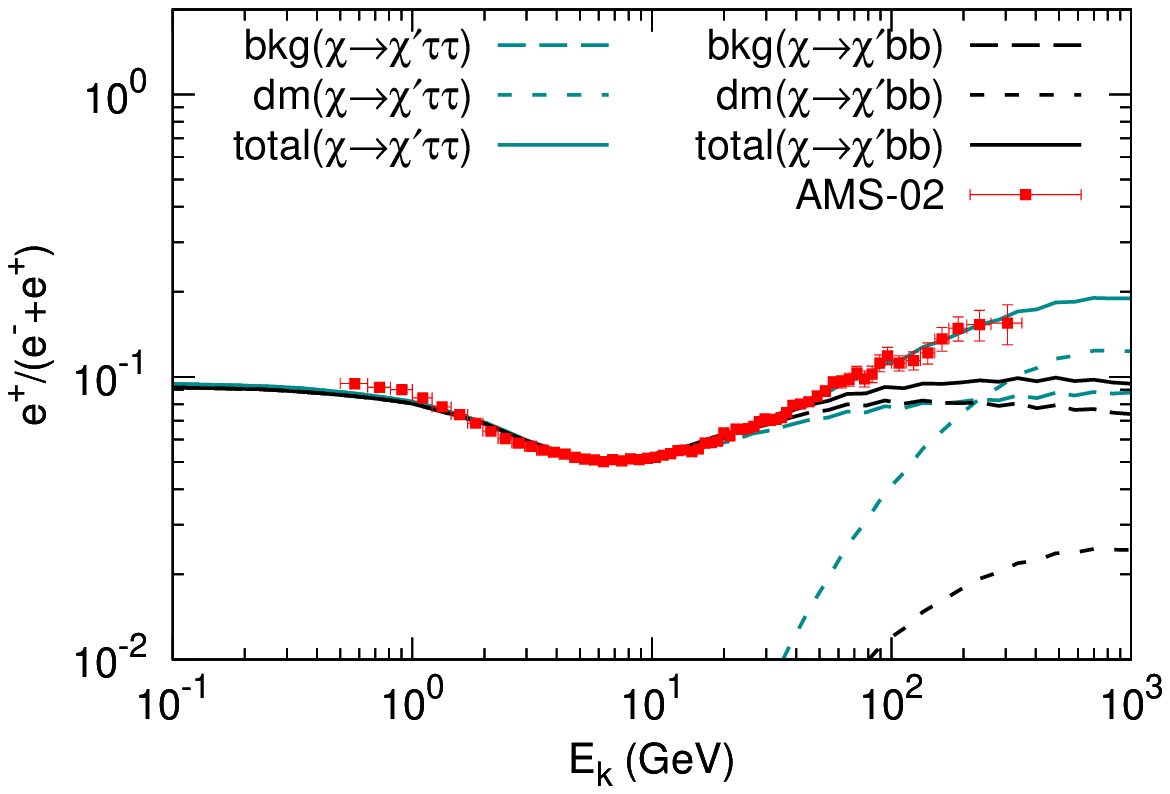}
\includegraphics[width=0.32\textwidth]{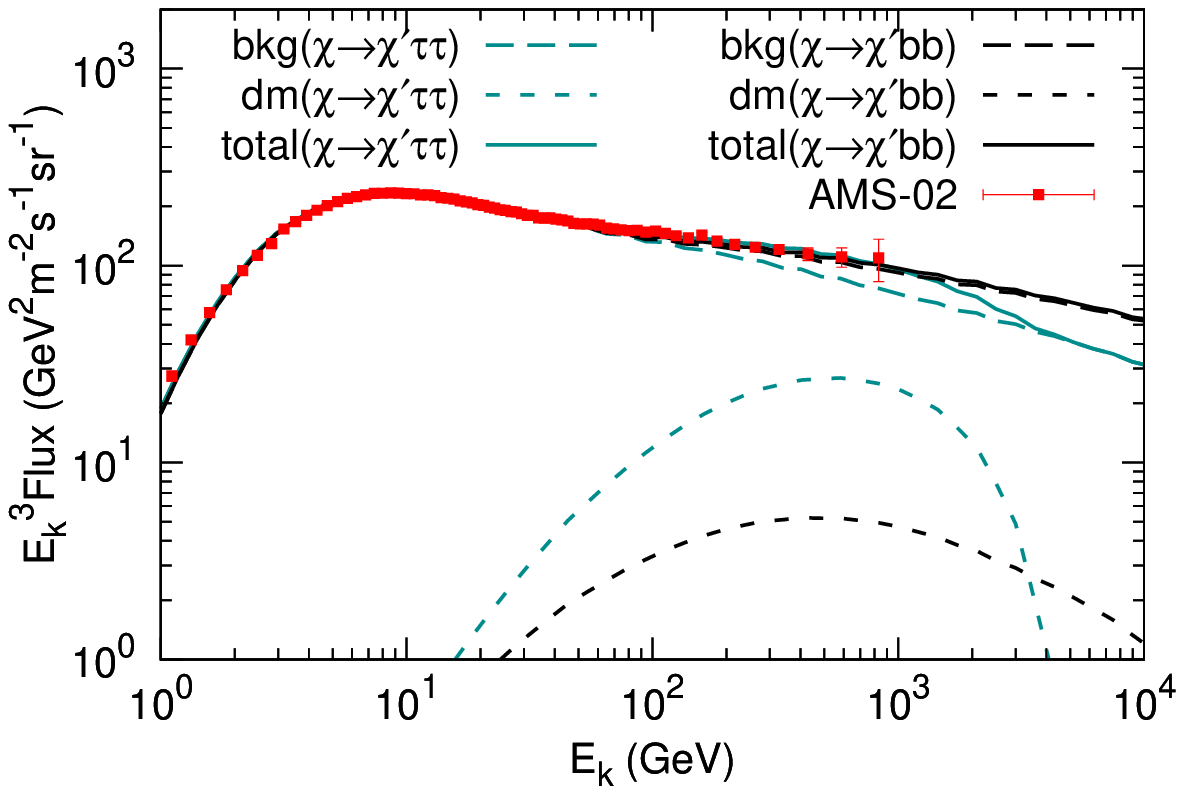}
\includegraphics[width=0.32\textwidth]{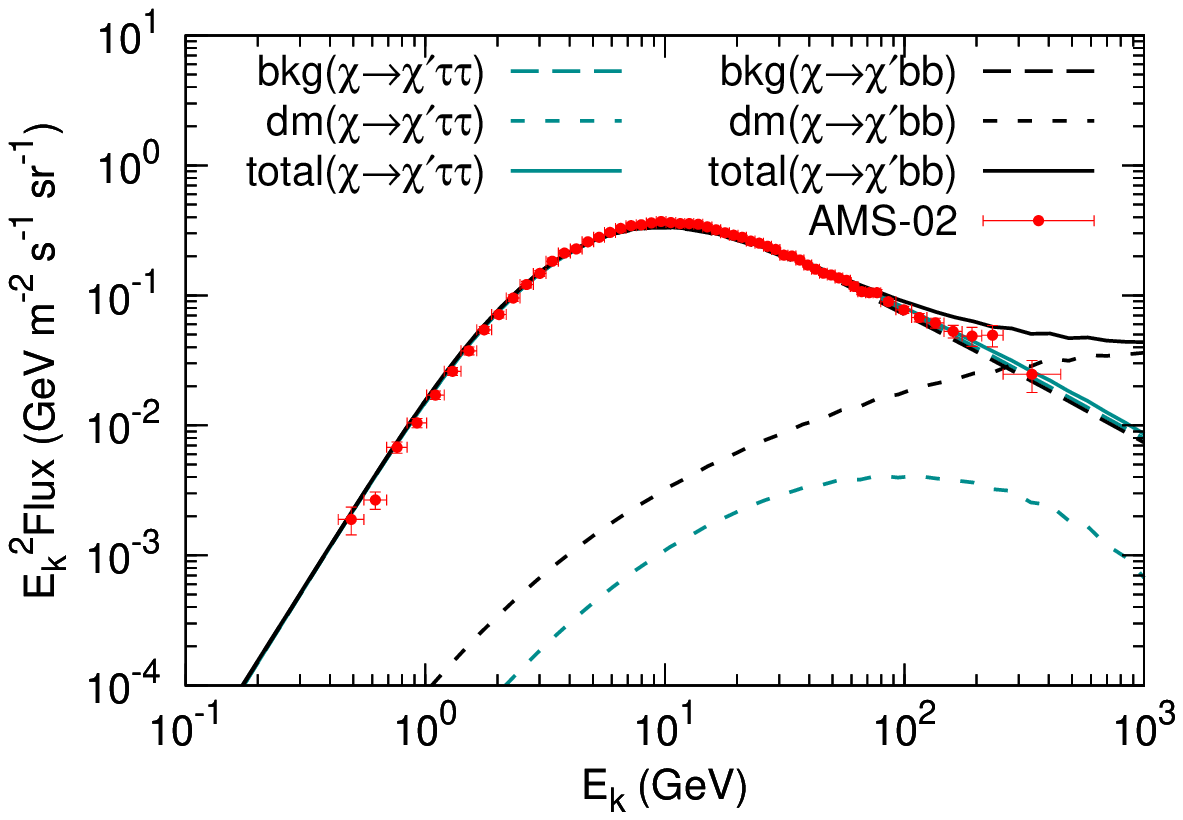}
\includegraphics[width=0.32\textwidth]{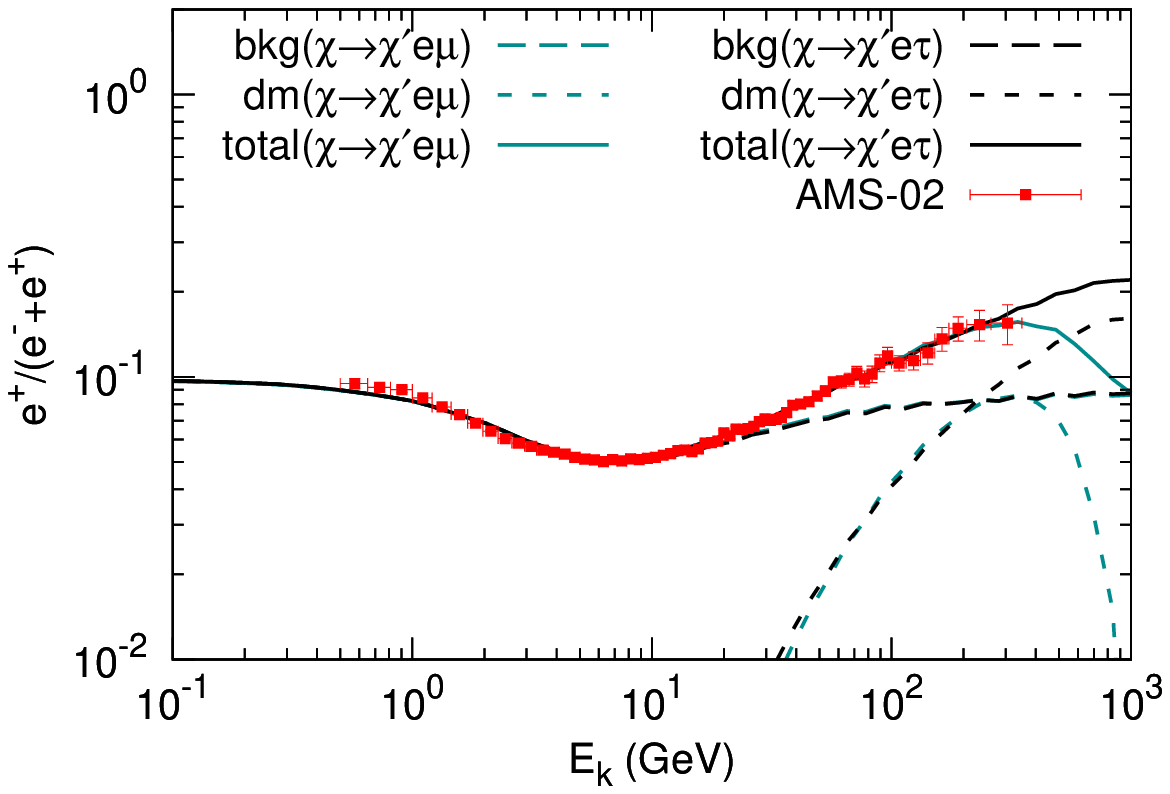}
\includegraphics[width=0.32\textwidth]{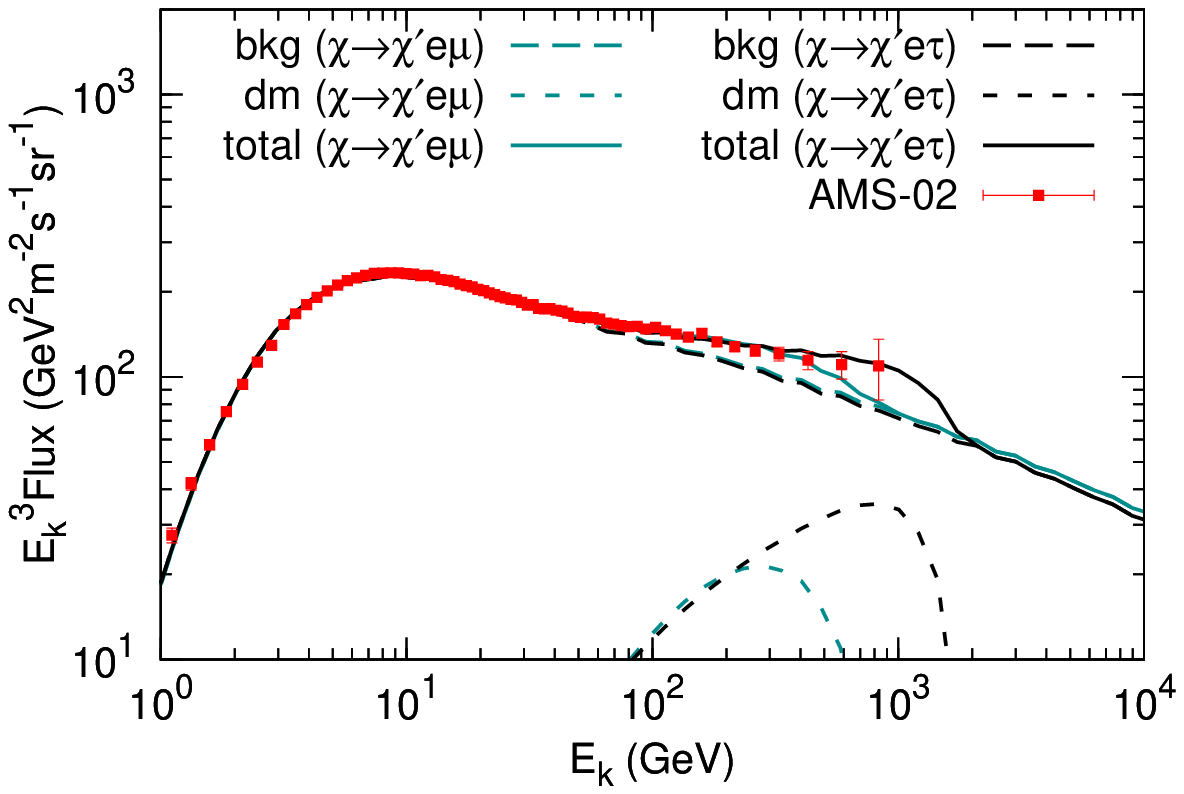}
\includegraphics[width=0.32\textwidth]{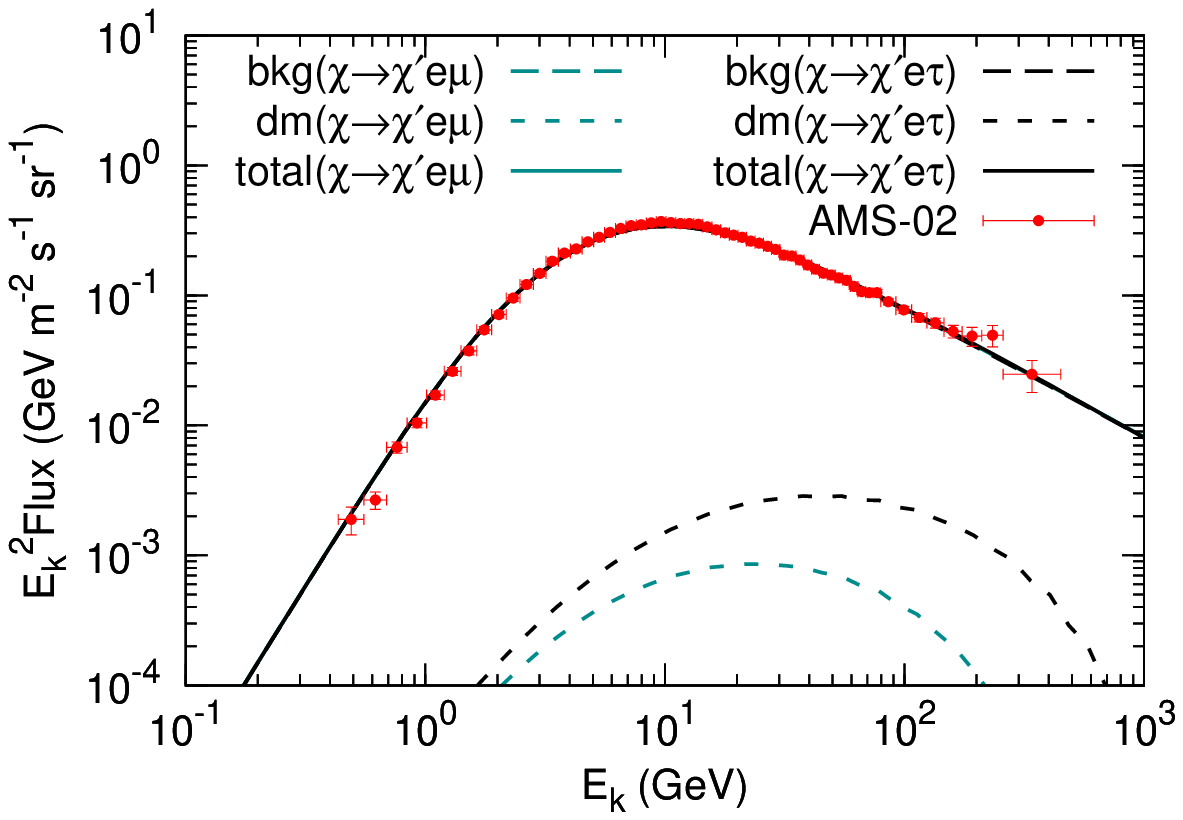}
\includegraphics[width=0.32\textwidth]{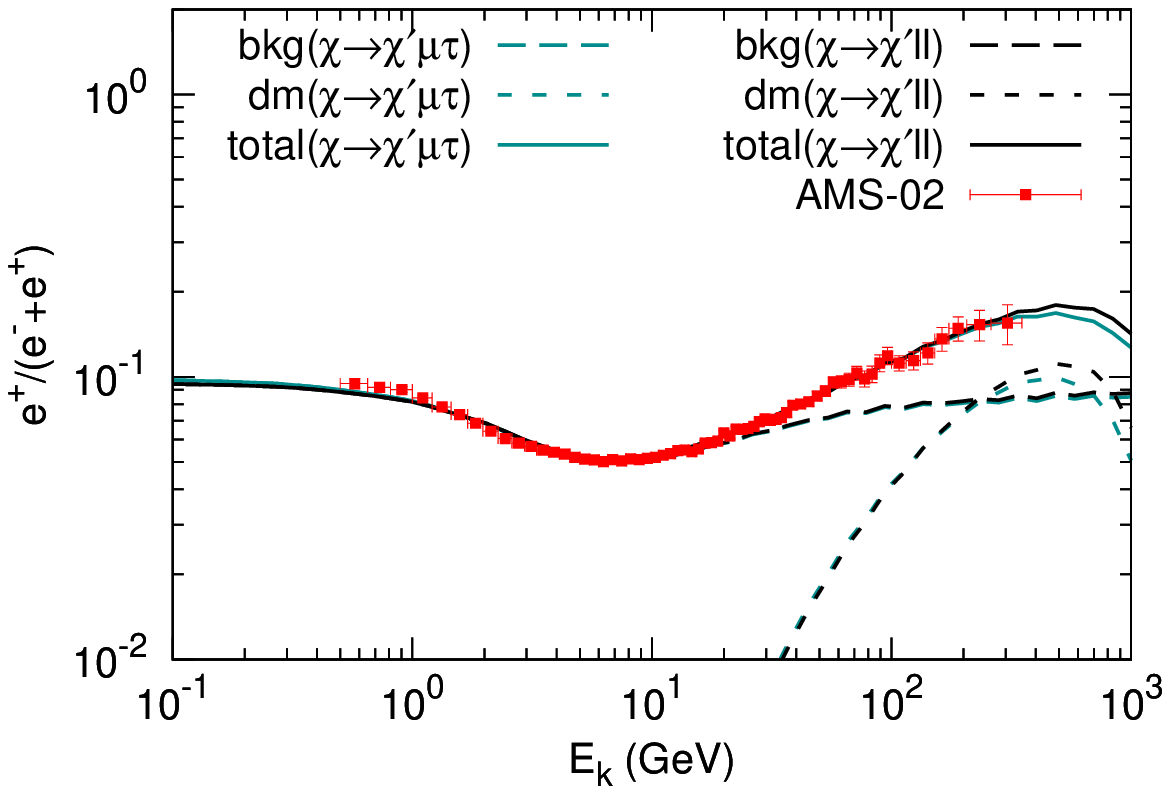}
\includegraphics[width=0.32\textwidth]{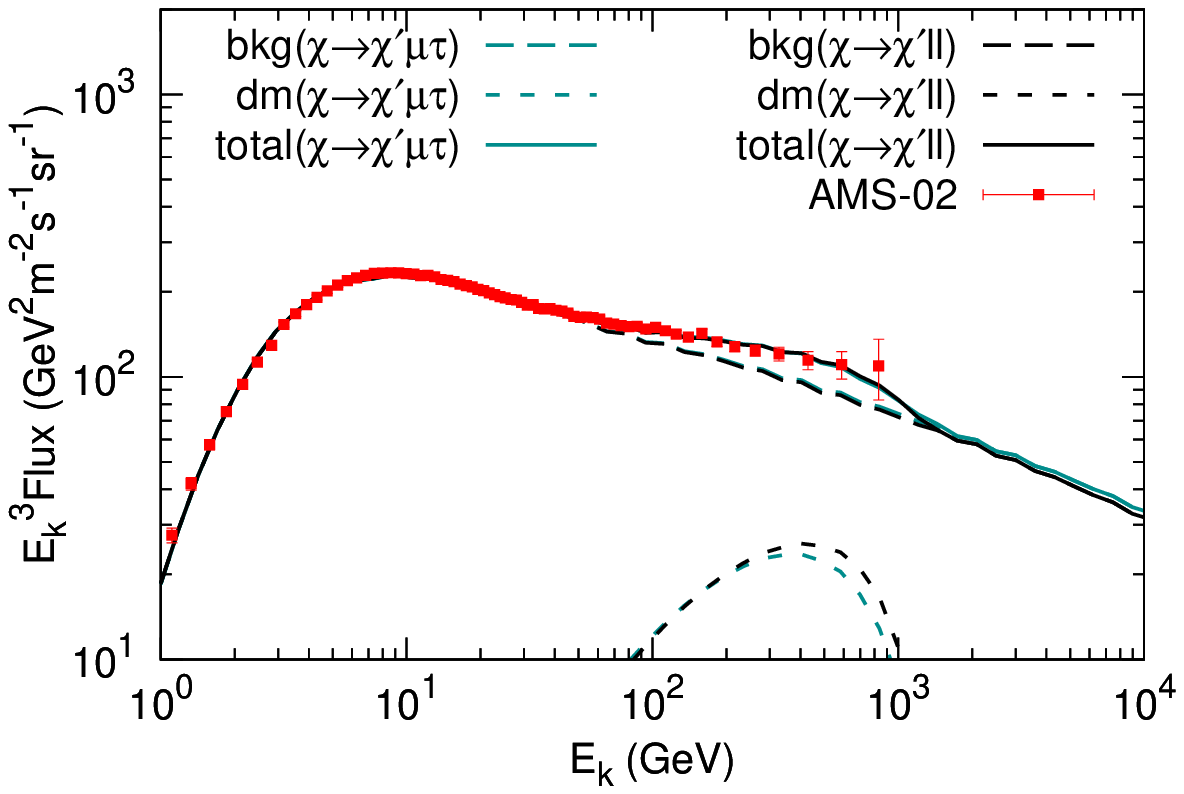}
\includegraphics[width=0.32\textwidth]{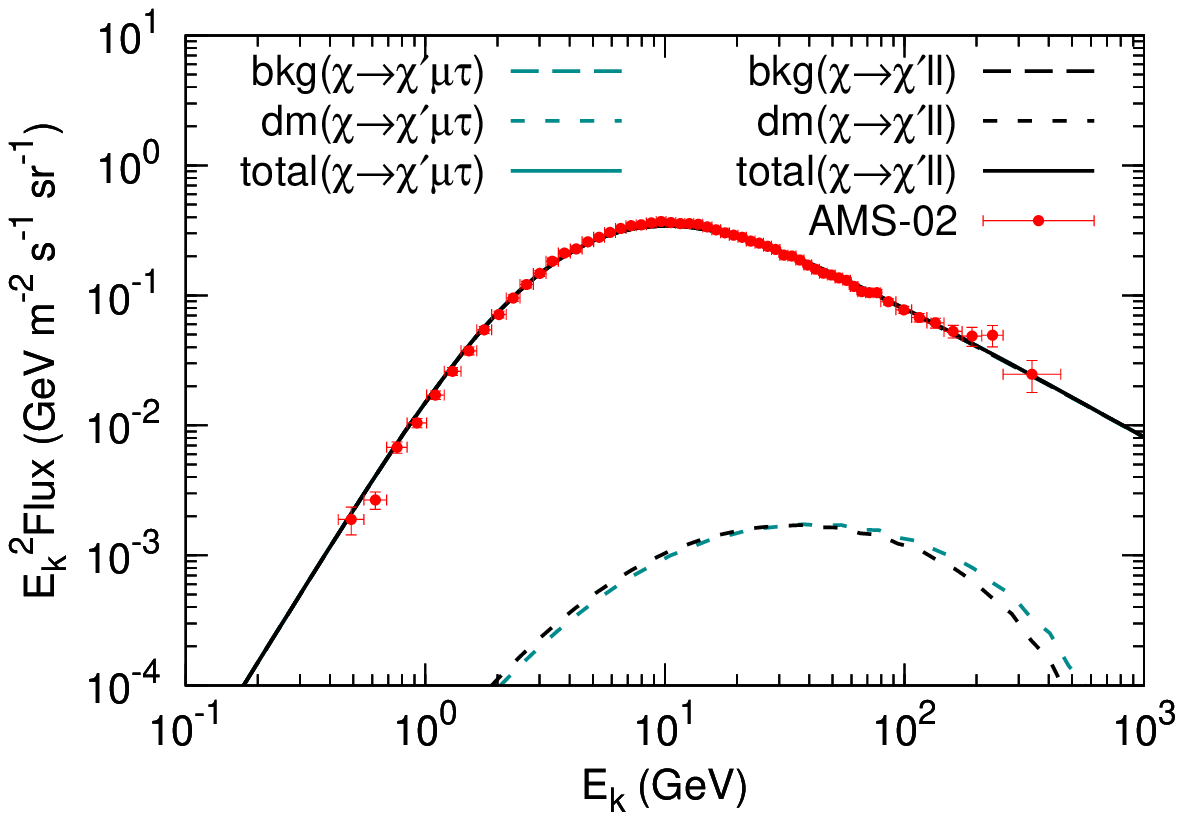}
\caption{Illustration of the best-fitting results for the positron fraction 
(left column), total $e^+e^-$ spectra (middle column), and anti-proton
spectra (right column), for the benchmark propagation parameters. 
The decay channels are labelled in the plot. The data from AMS-02 
(for the positron fraction and total $e^+e^-$ spectra
\cite{Accardo:2014lma,Aguilar:2014fea}, and anti-protons 
\cite{Aguilar:2016kjl}) are also shown.
}
\label{fig:AMS_ep_pb}
\end{figure}

Figure~\ref{fig:AMS_ep_pb} shows the comparisons of the positron fraction
(left column), total $e^+e^-$ spectra (middle column), and anti-proton 
spectra (right column) between the best-fitting model predictions and the 
measurements, for the benchmark propagation model. Eight decay channels, 
$\chi\to\chi'ee$, $\chi\to\chi'\mu\mu$,
$\chi\to\chi'\tau\tau$, $\chi\to\chi'b\bar{b}$, $\chi\to\chi'e\mu$,
$\chi\to\chi'e\tau$, $\chi\to\chi'\mu\tau$, and $\chi\to\chi'll$
($e:\mu:\tau=1:1:1$), are calculated. Except for the quark final state
which will over-produce anti-protons, the goodness-of-fittings of other
channels are comparable with each other. The reduced chi-squared values
of the fittings, $\chi^2_r$, are all about $1.0-1.1$ for 185 degrees of 
freedom except for $\chi\to\chi'b\bar{b}$ whose $\chi^2_r$ is about 3.0. 
The best-fitting parameters of $\Delta m_\chi$ and $m_{\chi}\tau$
are tabulated in Table \ref{table:para}. Note that 
the best-fitting values of $\Delta m_\chi$ in general increases as the 
decay channel varies from $e$, $\mu$, $\tau$, to $b$, from sub-TeV for 
the $e^+ e^-$ channel, multi-TeV for channels involving $\mu$ or $\tau$, 
to $100$ TeV for the $b\bar{b}$ channel\footnote{The channel 
$\chi\to\chi'e\tau$, however, has a larger $\Delta m_\chi$ than that of 
$\chi\to\chi' \mu\tau$ due to the fact that the likelihood distributions 
are a bit flat, as indicated by the elongated oval shape in credible regions.}.
In particular, the best-fit value of $\Delta m_\chi$ for the universal 
lepton decay channel $ll$ is 2.2 TeV, higher than $\sim 1$ TeV obtained 
previously in Refs.~\cite{Cheng:2010mw,Cheng:2012uk}, which were based on 
the Fermi-LAT $e^+ + e^-$ total flux~\cite{Abdo:2009zk} and the positron 
fraction data from PAMELA~\cite{Adriani:2008zr}. This is in large part 
due to the updated electron background, besides fitting to the newer 
AMS-02 data. As a result it will produce more high energy $\gamma$-rays, 
and as we will see in the next section, the universal lepton decay channel 
is now disfavored. 

\begin{table}[!htb]
\centering
\caption{Best-fitting DM parameters}
\begin{tabular}{cccccccccc}
\hline \hline
 & Unit & $ee$ & $\mu\mu$ & $\tau\tau$ & $b\bar{b}$ & $e\mu$ & $e\tau$ & $\mu\tau$ & $ll$\\
\hline
$\Delta m_\chi$ & (TeV) & 0.6 & 2.0 & 8.3 & 100.0$^\dagger$ & 1.2 & 3.0 & 2.8 & 2.2 \\
$m_{\chi}\tau$  & ($10^{27}$TeV s) & 0.4 & 1.3 & 1.4 & 57.5 & 0.8 & 0.7 & 1.1 & 0.8 \\
\hline
\hline
\end{tabular}\vspace{3mm}\\
$^\dagger$This value is close to the upper bound of the scan.\\
\label{table:para}
\end{table}

\begin{figure}[!htb!]
\centering
\includegraphics[width=0.43\textwidth]{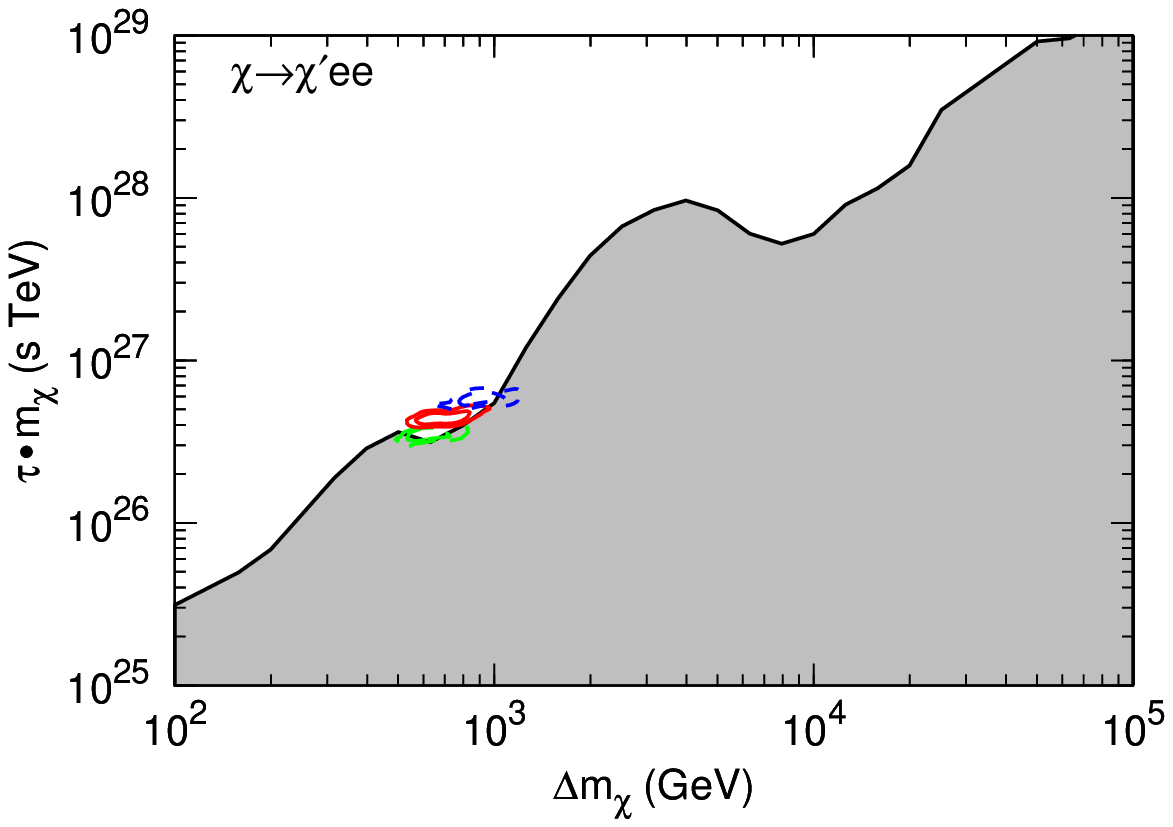}
\includegraphics[width=0.43\textwidth]{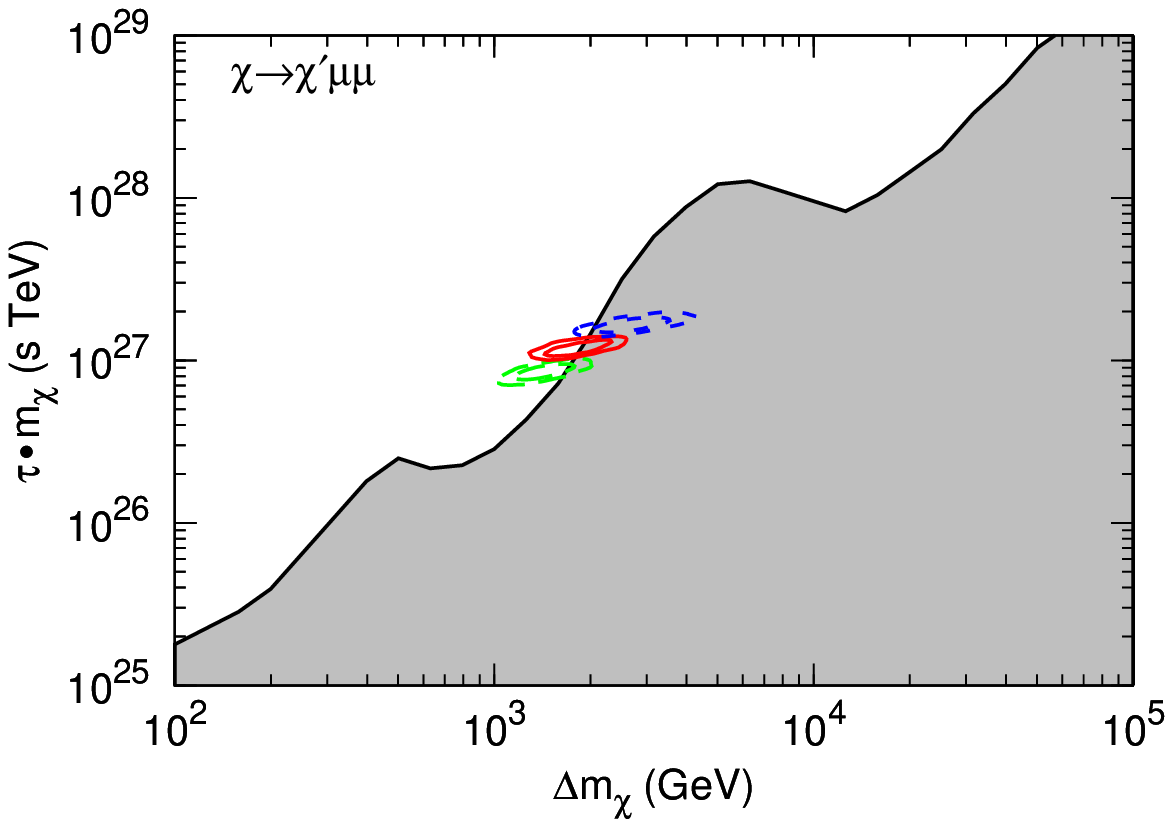}
\includegraphics[width=0.43\textwidth]{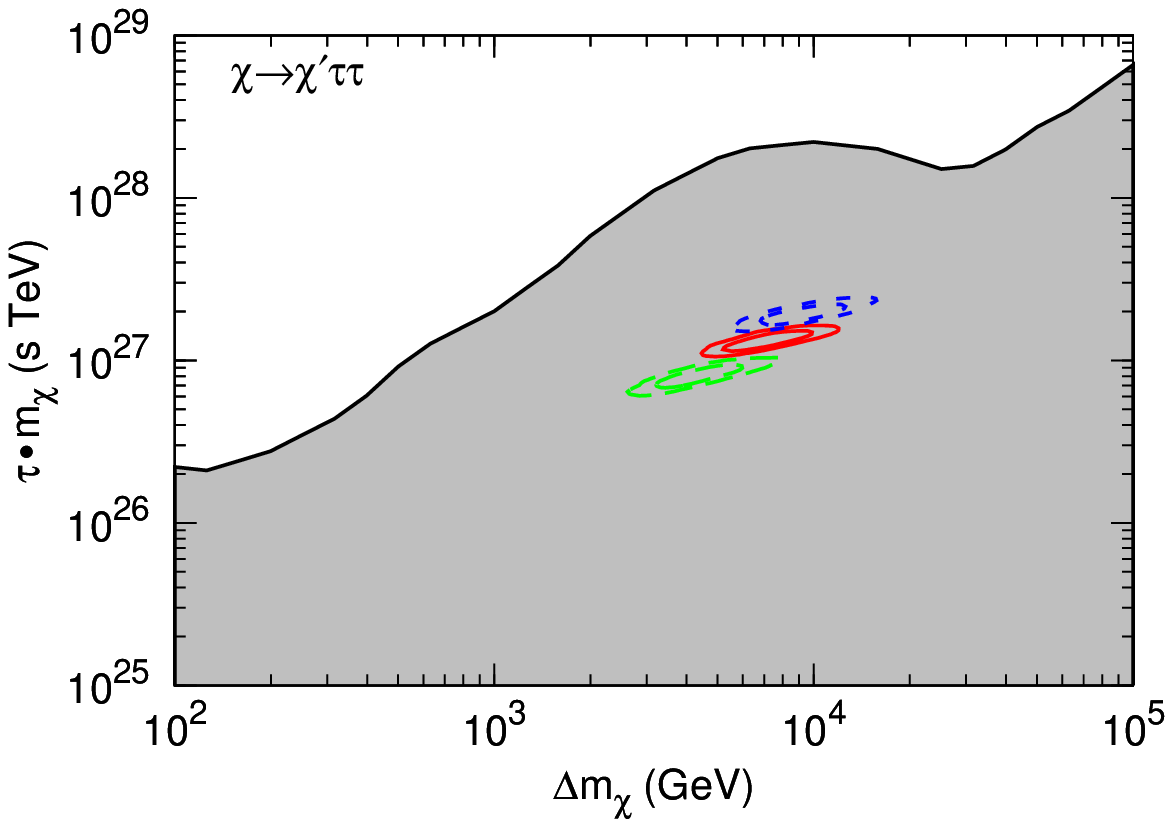}
\includegraphics[width=0.43\textwidth]{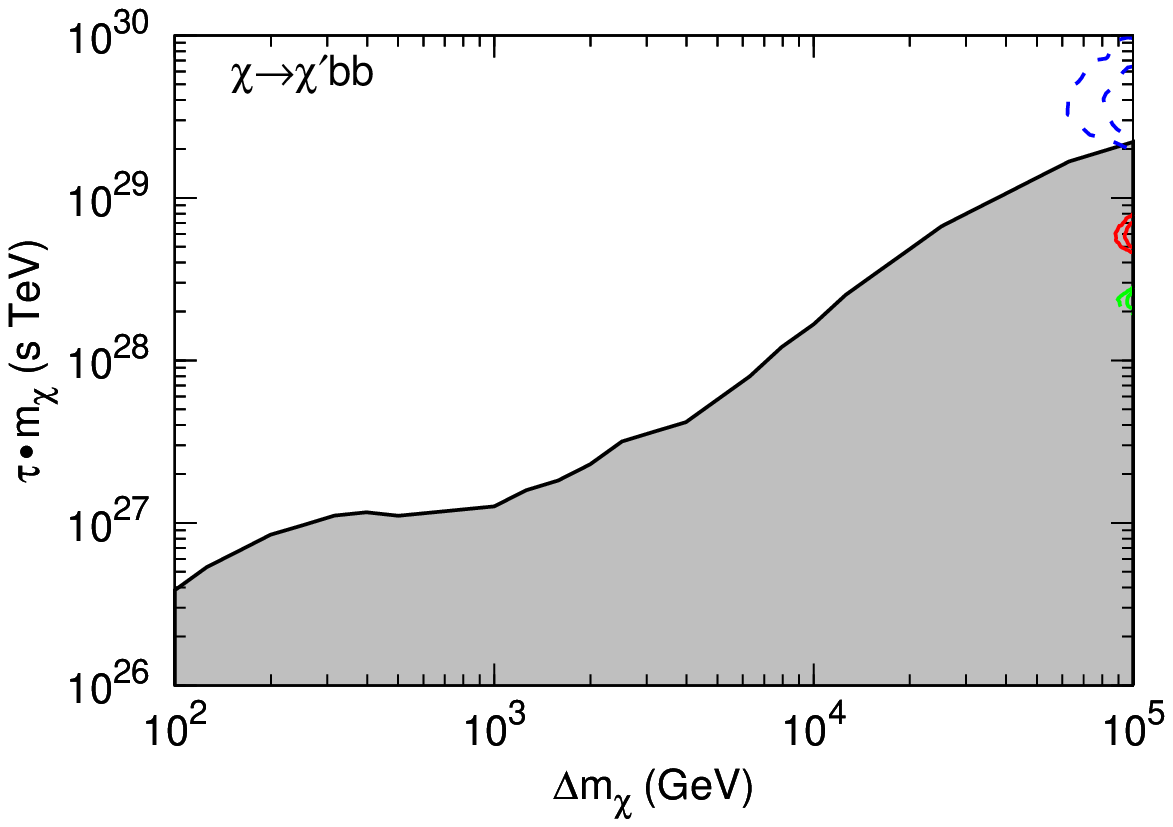}
\includegraphics[width=0.43\textwidth]{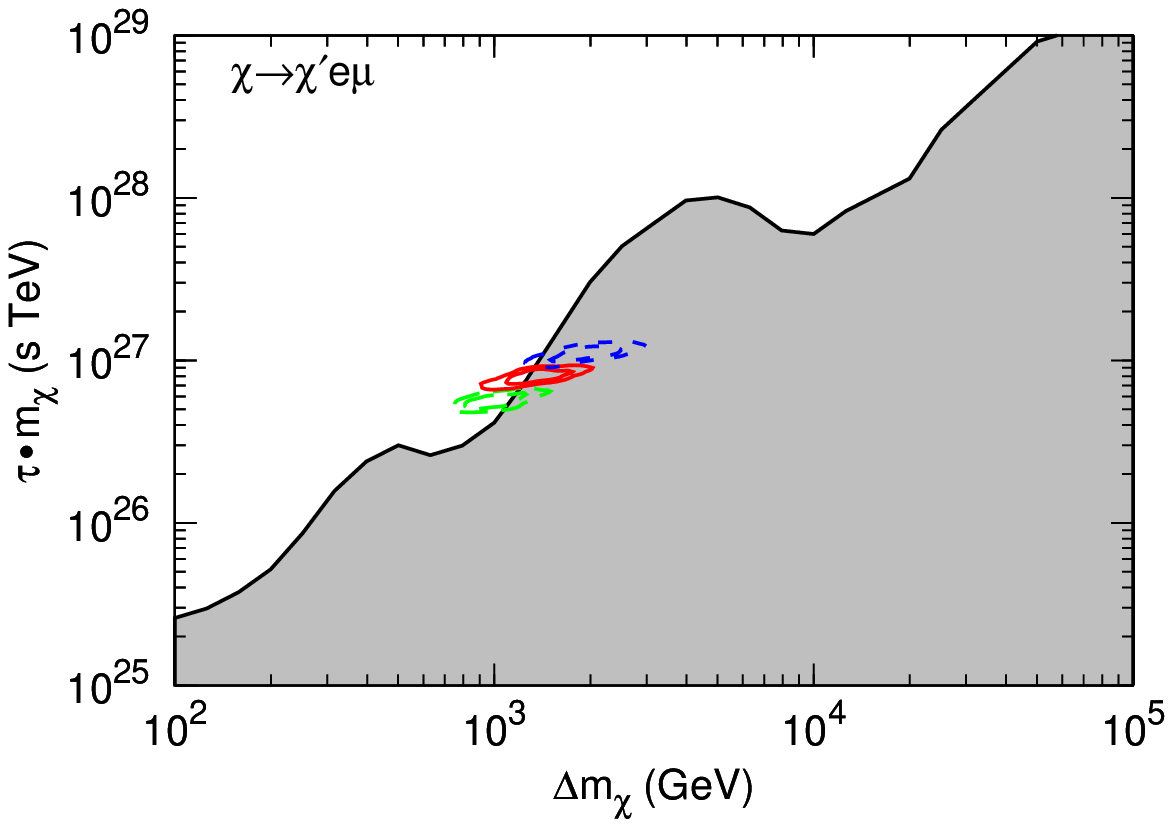}
\includegraphics[width=0.43\textwidth]{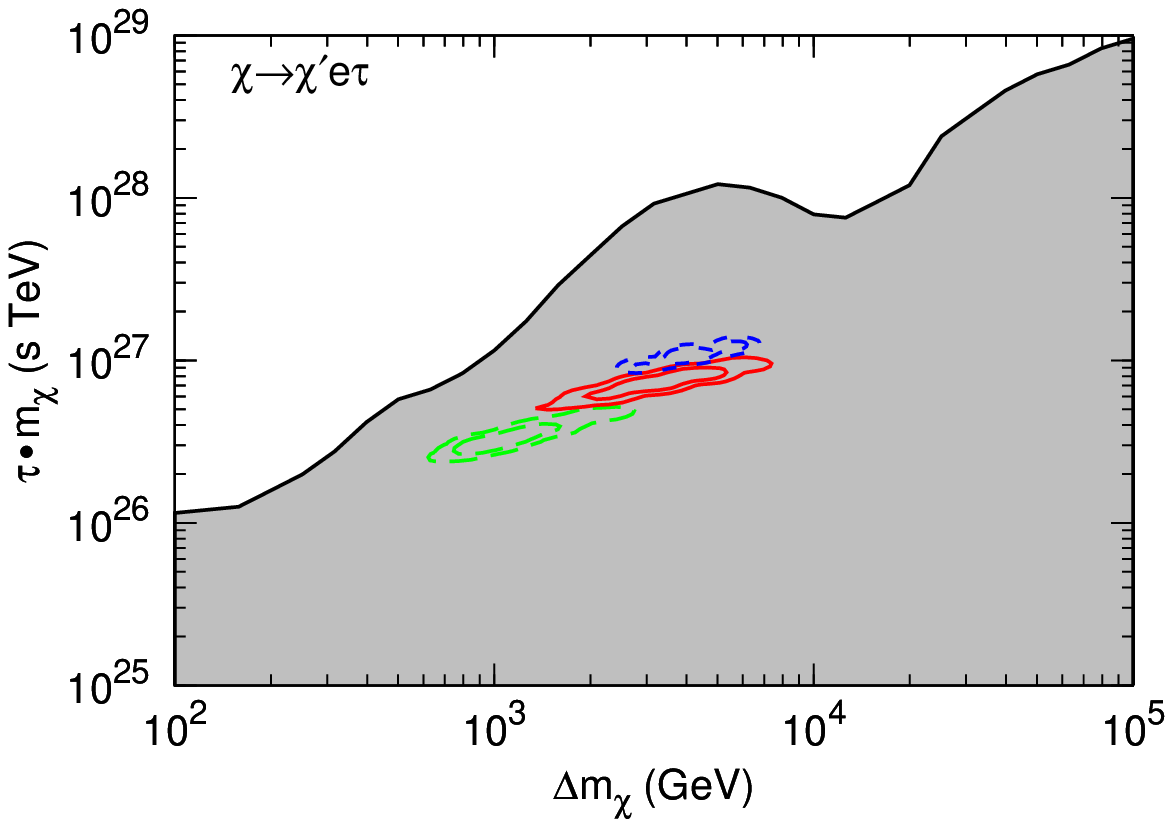}
\includegraphics[width=0.43\textwidth]{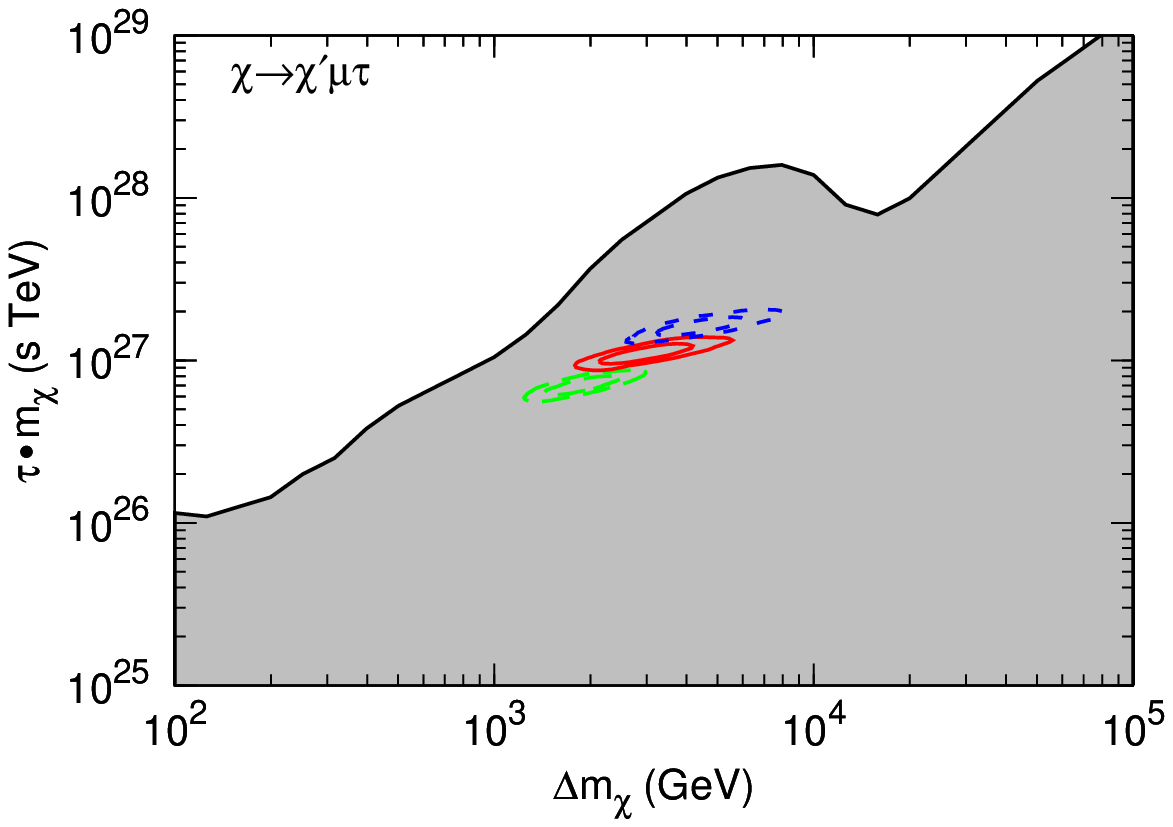}
\includegraphics[width=0.43\textwidth]{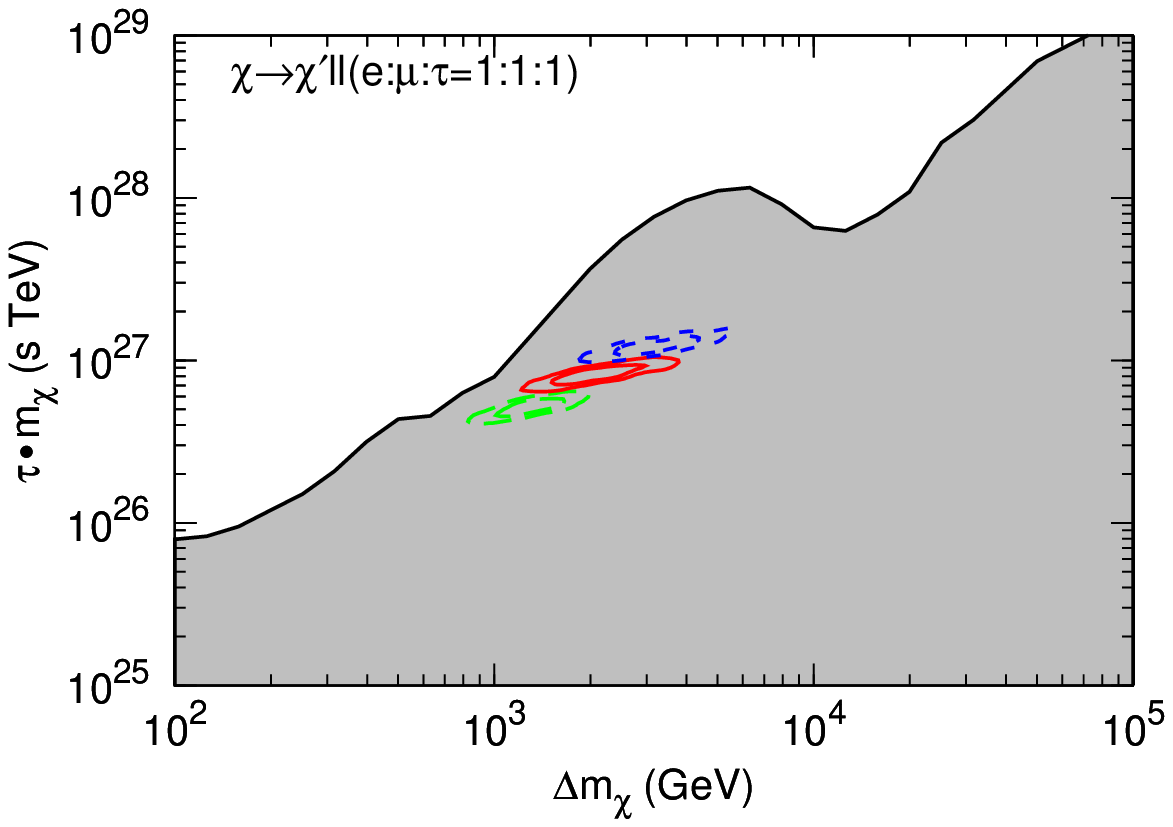}
\caption{Contours of $68\%$ (inner) and $95\%$ (outer) credible regions
on the $\Delta m_{\chi}-m_{\chi}\tau$ plane, from the fittings of the 
charged CR data. The green, red, and blue contours correspond to the 
propagation parameter settings 1, 2, and 6, as shown in Table 
\ref{table:prop}. The shaded regions are disfavored by the Fermi EGB 
data at the $95\%$ credible interval (see Sec. 4).} 
\label{fig:EGcons}
\end{figure}
    
The fitting contours on the parameter plane of $\Delta m_{\chi}$ and
$m_{\chi}\tau$, for the eight decay channels, are shown in 
Figure~\ref{fig:EGcons}. In each panel, the green, red, and blue contours
are for propagation parameter setting 1, 2, and 6, respectively, which 
are shown in Table \ref{table:prop}. As shown in Ref. \cite{Yuan:2014pka}, 
such three parameter settings are typical to represent the systematic 
uncertainties from the CR propagation, especially for this high energy 
region. As an illustration, we show the posterior probability distributions 
of all parameters and their correlations for the case of the $ee$ channel 
in the Appendix.

\section{Constraints from EGB}

In this section we discuss the compatibility of the DM model to explain
the AMS-02 electron/positron data with the point source subtracted EGB 
as measured by Fermi-LAT \cite{Ackermann:2014usa}. The diffuse $\gamma$-ray 
emission from extragalactic DM decay is (e.g., \cite{Chen:2009uq})
\begin{equation}
\phi_{\rm EG}(E)=\frac{c}{4\pi}\frac{\Omega_{\chi}\rho_c}{m_{\chi}\tau}
\int_0^{\infty}\frac{dz'}{H(z')}\frac{dN}{dE'}\exp[-\tau(z',E')],
\label{eq:phi_eg}
\end{equation}
where $\Omega_\chi\rho_c$ is the current DM density in the Universe,
$H(z)=H_0\sqrt{\Omega_M(1+z)^3+\Omega_{\Lambda}}$ is the Hubble parameter,
$E'=E(1+z)$, $\frac{dN}{dE'}=\left.\frac{dN}{dE'}\right|_{\rm prompt}+
\left.\frac{dN}{dE'}\right|_{\rm IC}$ is the $\gamma$-ray spectrum yielded at redshift $z'$ for one decay of a DM particle, and $\tau(z,E)$ is the
attenuation optical depth of high energy $\gamma$-rays in the cosmic
infrared to ultraviolate background. In this work we use the extragalactic
background light model of Ref. \cite{Gilmore:2011ks}. The $\gamma$-ray 
spectrum consists of two parts, the prompt emission associated with 
the decay of DM particles and the IC emission from the decay products 
$e^+e^-$. We use the same way as that of electrons/positrons to calculate 
the prompt $\gamma$-ray emission from three-body DM decay 
(see Eq.~(\ref{eq:secondary})). For the IC emission, we assume instantaneous 
cooling of $e^+e^-$ after their production in the cosmic microwave background 
(CMB), which gives the equilibrium $e^+e^-$ spectrum (per 
decay\footnote{To obtain the energy spectrum per unit volume, one should 
multiply it by the decay rate, $\Omega_{\chi}\rho_c(1+z)^3/(m_{\chi}\tau)$. 
Here we keep this form in order to calculate the IC emission in the same 
way as that of the prompt emission by Eq. (\ref{eq:phi_eg}).}) as 
\cite{Liu:2016ngs}
\begin{equation}
\frac{d{\mathcal N}}{dE_e}(E_e,z)=\frac{1}{b(E_e,z)}\int_{E_e}^{\infty}
dE_e'\frac{dN}{dE_e'},
\end{equation}
where $b(E_e,z)=2.67\times 10^{-17}(1+z)^4(E_e/{\rm GeV})^2$ GeV s$^{-1}$
is the IC energy loss rate. The IC photon spectrum can then be calculated
using the Klein-Nishina differential scattering cross section 
\cite{Blumenthal:1970gc}. The cosmological parameters used are: $H_0=67$ 
km s$^{-1}$ Mpc$^{-1}$, $\Omega_M=0.32$, $\Omega_\Lambda=0.68$, 
$\Omega_\chi=0.27$, and $\rho_c=1.24\times10^{11}$ M$_{\odot}$ 
Mpc$^{-3}$ \cite{Ade:2013zuv}. 

The minimum emission from Galactic DM decay (from the anti-GC direction) 
will also contribute to the EGB \cite{Cirelli:2012ut}. It can be calculated as
\begin{equation}
\phi_{\rm G}(E)|_{{\rm anti-GC}}=\frac{1}{4\pi m_{\chi}\tau}\frac{dN}{dE}
\times \int_{\rm l.o.s.} \rho(l)dl,
\label{anti-GC}
\end{equation}
in which the integration is taken along the line-of-sight toward the anti-GC
direction. Again $dN/dE$ consists of the prompt and the IC components.
Since most of the DM particles from the anti-GC direction locate outside the 
propagation halo of Galactic CRs, we neglect the diffusion and consider 
only the IC cooling in the CMB of $e^+e^-$. So the calculation of 
the IC emission is identical to the extragalactic case described above.
Note that the method to calculate the IC emission from the GC direction 
as will be discussed in Sec.~\ref{sec:GC} is different, where the diffusion of
$e^+e^-$, as well as the cooling in the Galactic optical/infrared
field and magnetic field, needs to be included.

\begin{figure}[!htb!]
\centering
\includegraphics[width=0.45\textwidth]{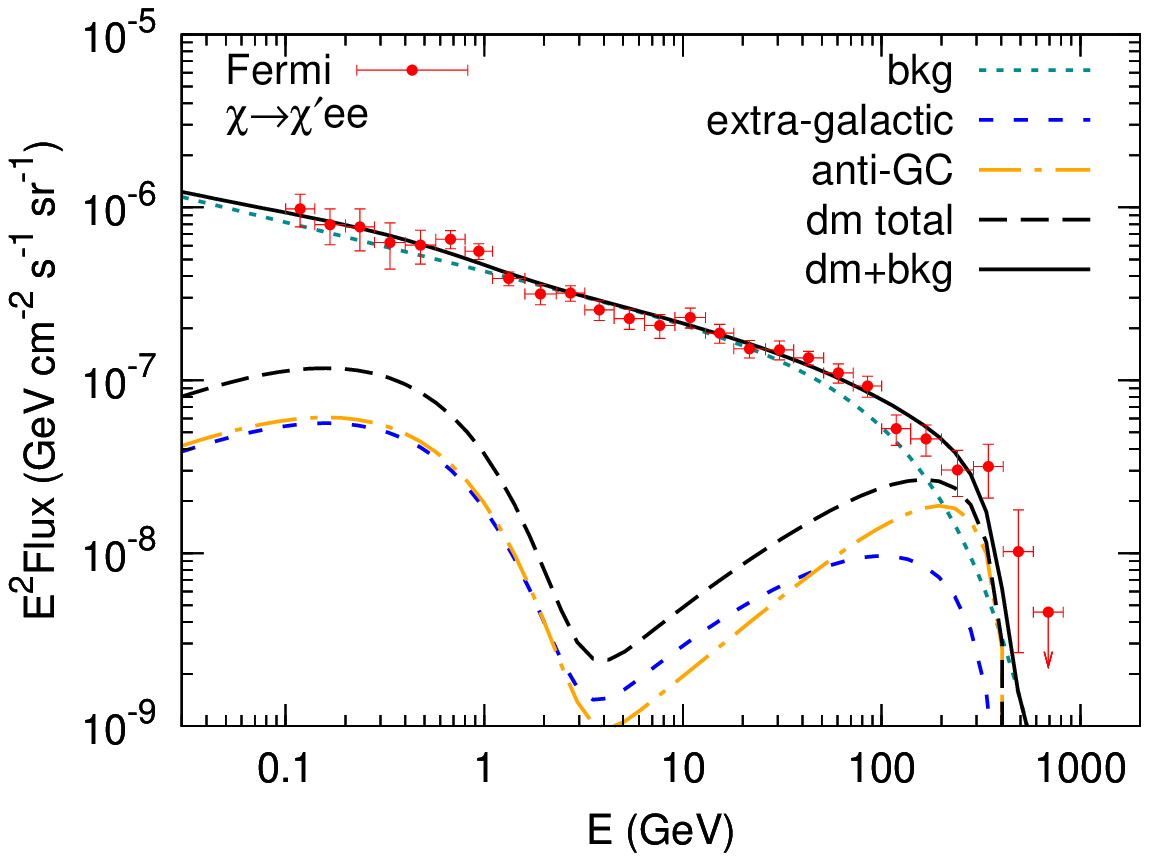}
\includegraphics[width=0.45\textwidth]{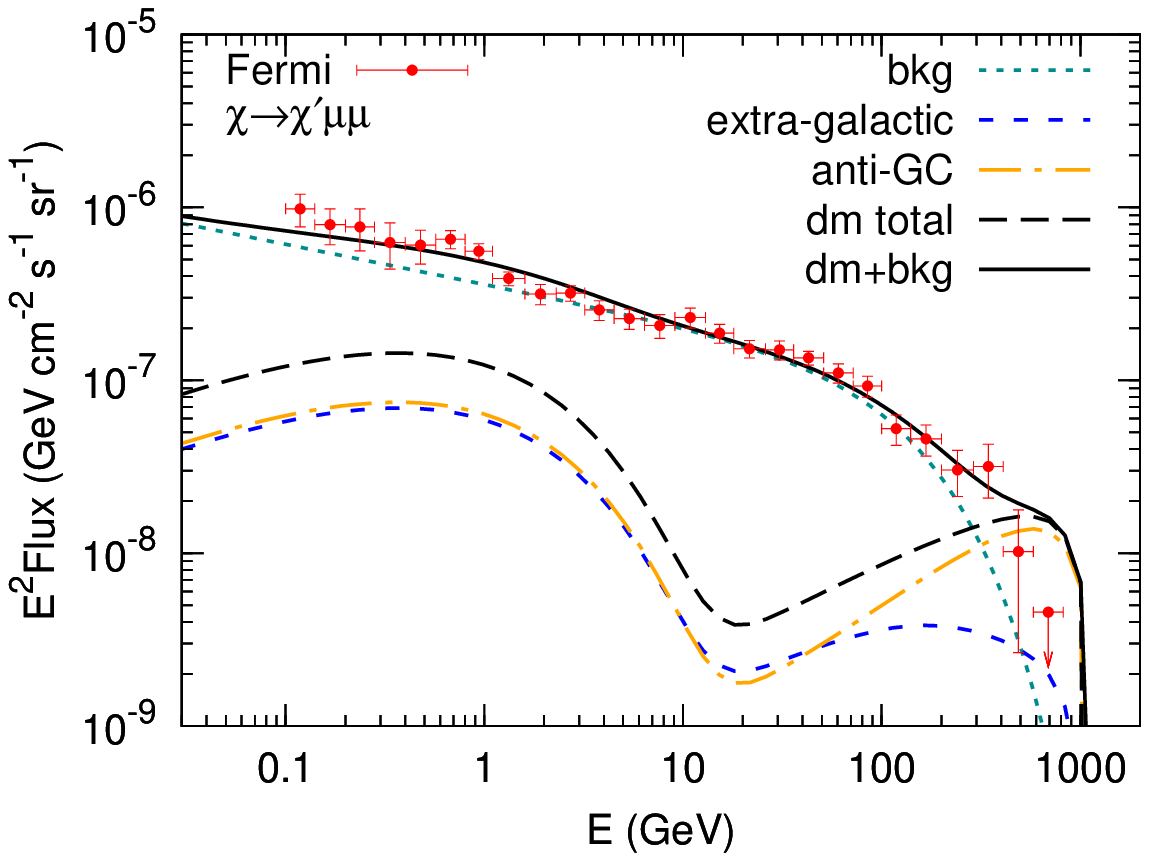}
\includegraphics[width=0.45\textwidth]{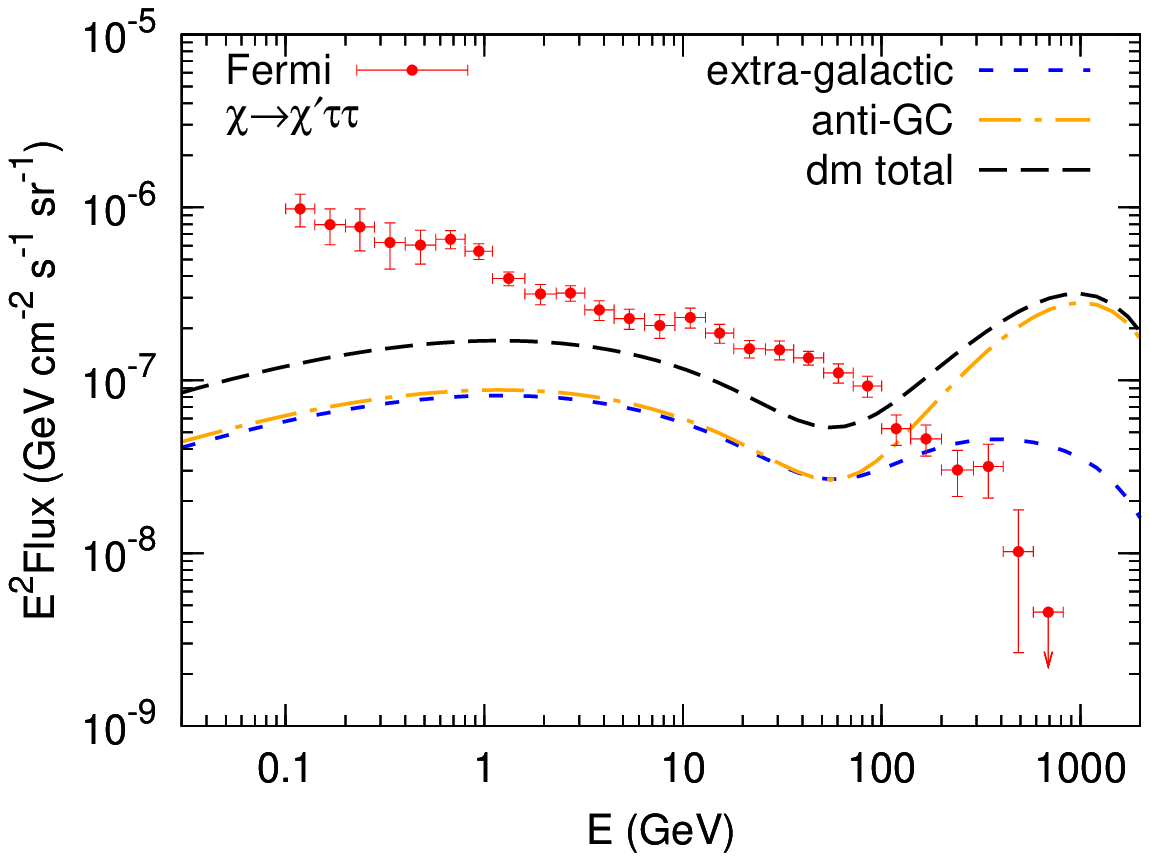}
\includegraphics[width=0.45\textwidth]{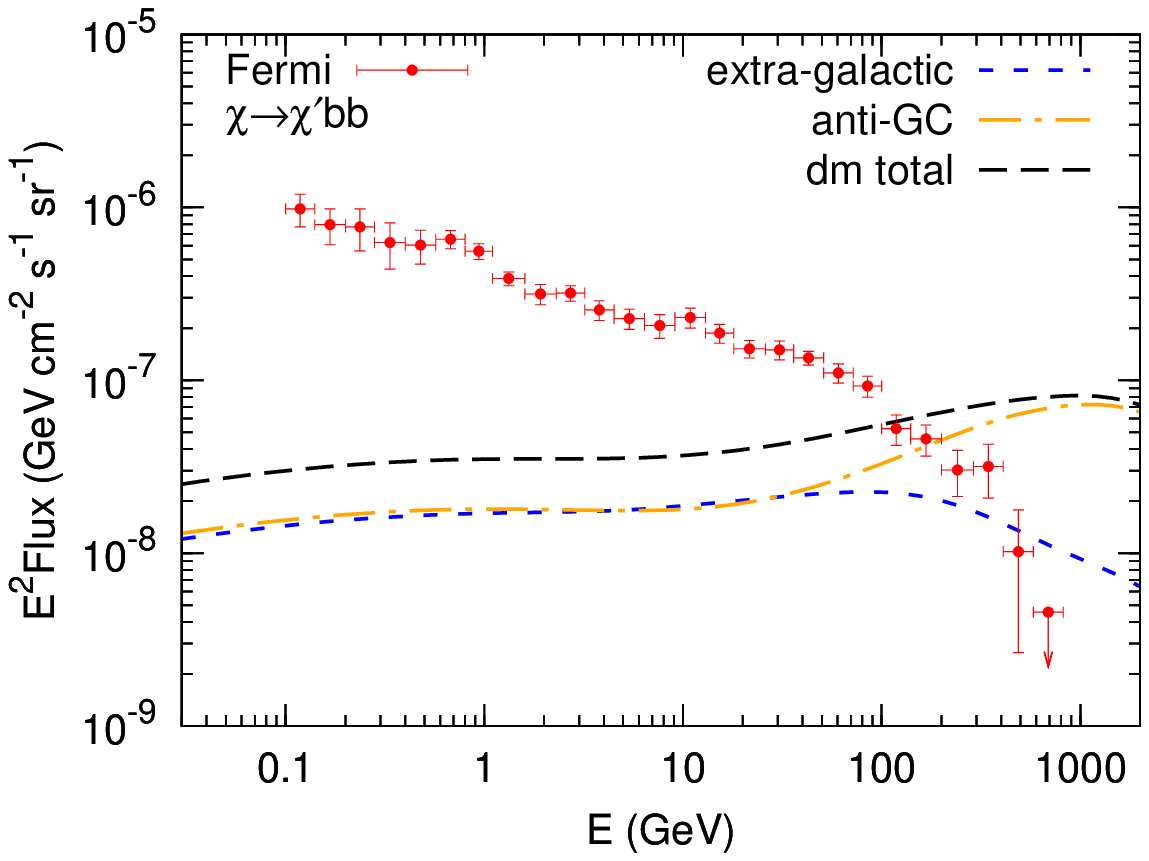}
\includegraphics[width=0.45\textwidth]{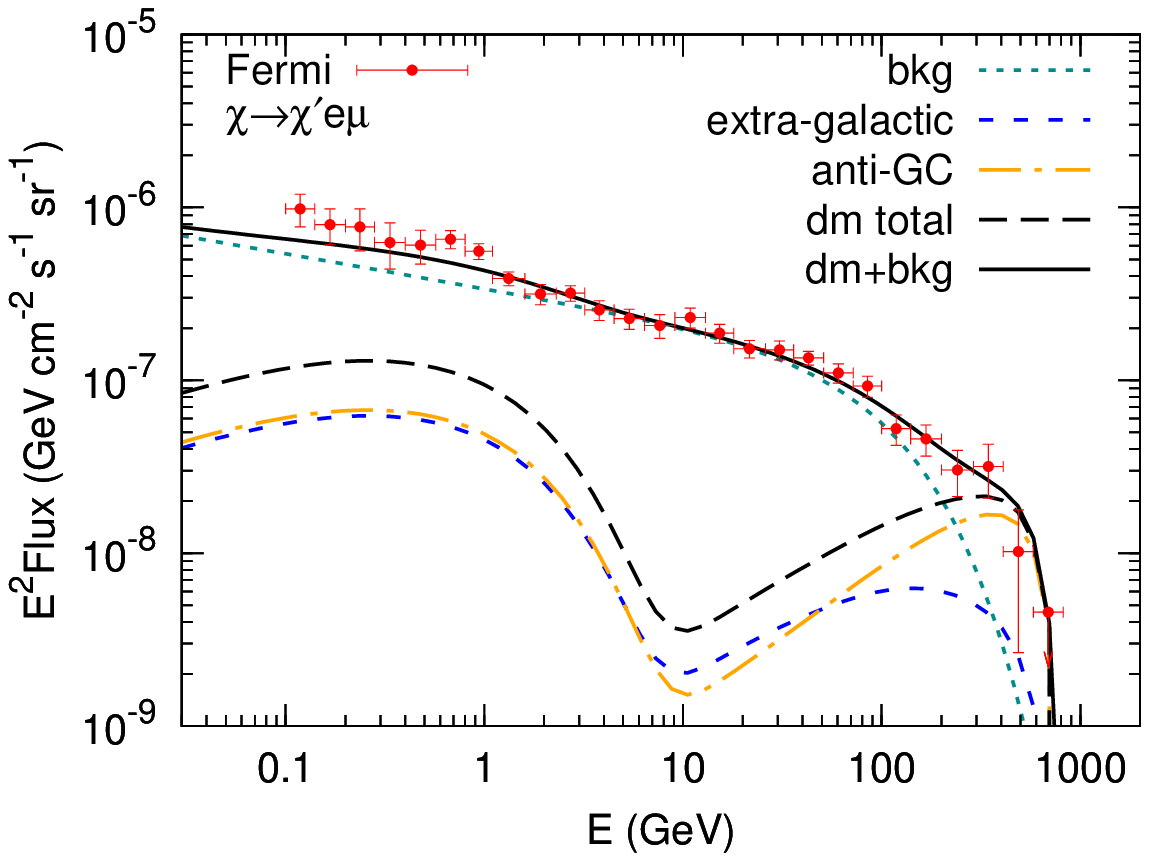}
\includegraphics[width=0.45\textwidth]{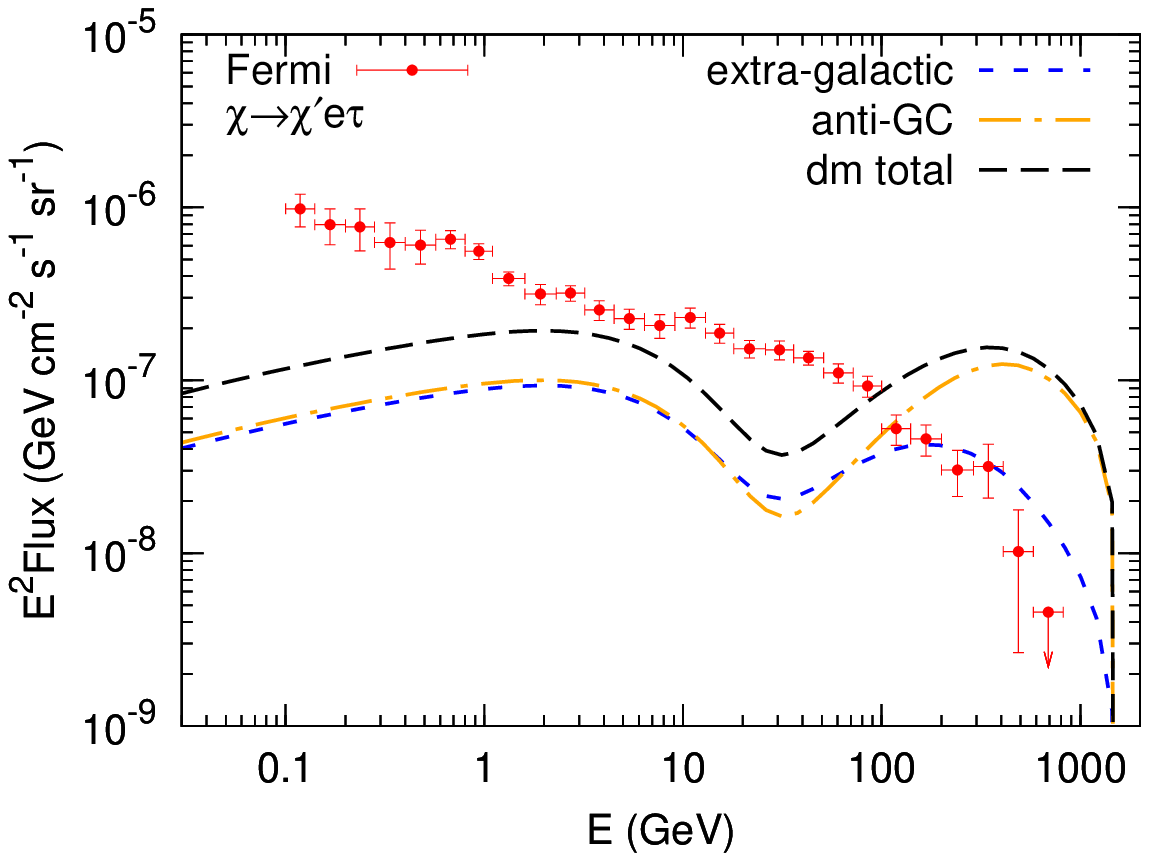}
\includegraphics[width=0.45\textwidth]{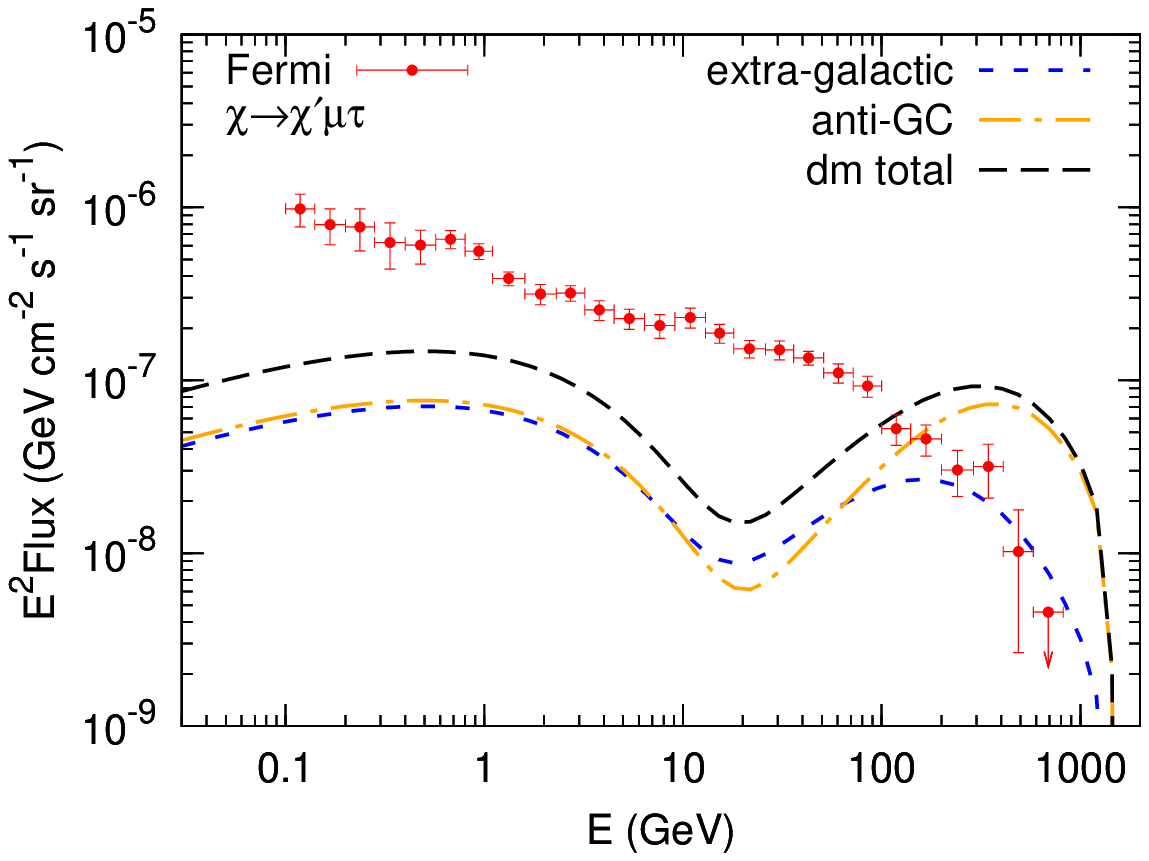}
\includegraphics[width=0.45\textwidth]{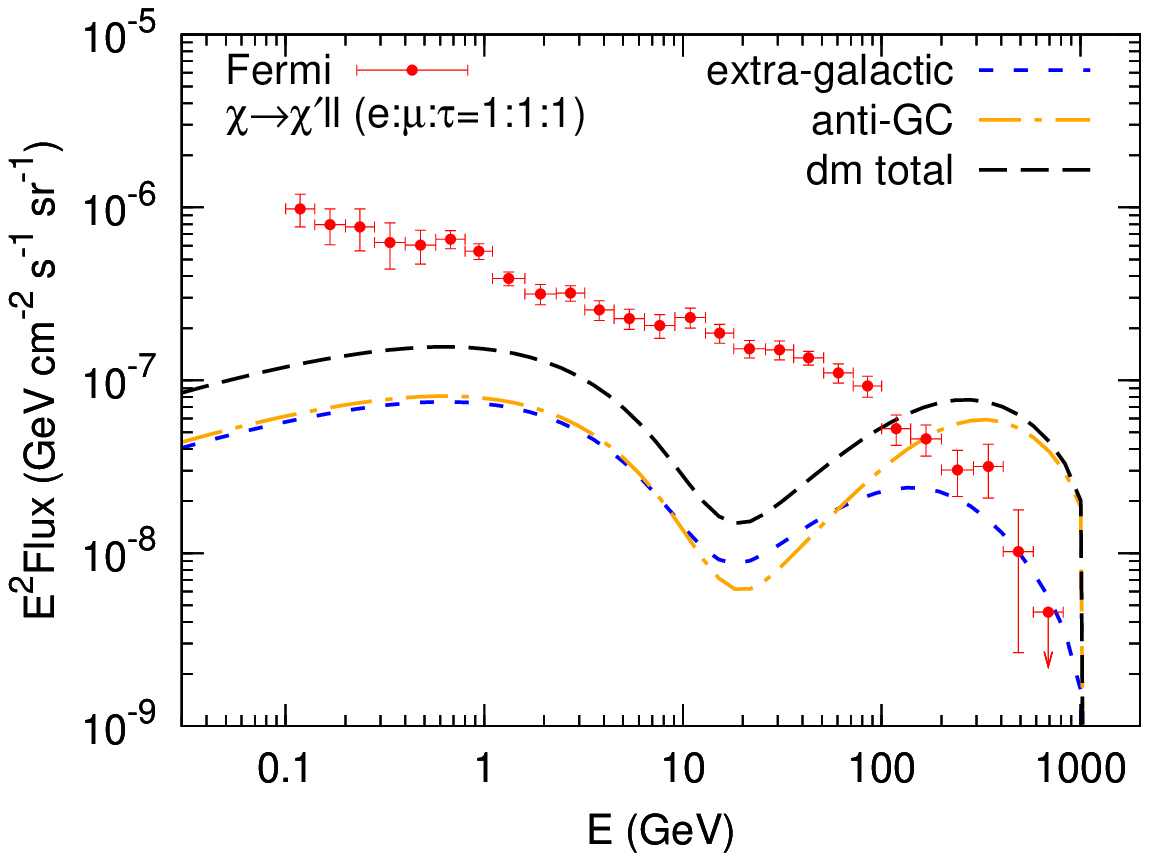}
\caption{Expected $\gamma$-ray emission from the best-fit DM models, 
compared with the EGB data of Fermi-LAT \cite{Ackermann:2014usa}.
}
\label{fig:EG}
\end{figure}

Figure~\ref{fig:EG} shows the expected diffuse $\gamma$-ray emission
from the decay of DM from extragalactic space (blue dashed) and the 
anti-GC direction (red dash-dotted), for the best-fit models as
shown in Figure~\ref{fig:AMS_ep_pb}. The two bumps of each component
correspond to the prompt (higher energy) and IC (lower energy) emissions.
We  find that the prompt emission of the extragalactic component
is suppressed by a factor of a few at the high energies due to the
attenuation in the extragalactic background light field. From this
plot we can see that if there are $\tau$ leptons or quarks in the decay
final state, the DM contribution will exceed the observational data
at high energies. Therefore, only the case with $ee$, $\mu\mu$, or
$e\mu$ final state particles can potentially survive from the data.

The measured spectrum of the EGB can be well described by an exponential 
cut-off power-law function, which is expected to come from point source
populations such as blazars \cite{TheFermi-LAT:2015ykq}. Fitting to the
data with only the background gives a power-law index of $\gamma=2.32\pm0.01$ 
and a cutoff energy of $E_{\rm cut}=288\pm41$ GeV. Such a spectrum 
is very distinct from the two-bump structure of the photon spectrum from 
the DM decay. Therefore we assume an exponential cut-off power-law 
function as the background and include the DM contribution in the fit to the data. The green dotted line (for $\chi\to\chi'ee$, 
$\chi\to\chi'\mu\mu$, and $\chi\to\chi'e\mu$ only) shows the curve 
that best-fits the data when fixing the DM model parameters as those 
in Table \ref{table:para}. 

We can derive constraints on the DM model parameters using the EGB data. 
The posterior probability density of the parameter $m_{\chi}\tau$ for any 
given $\Delta m_{\chi}$ can be written as 
\begin{equation}
{\mathcal P}(m_{\chi}\tau)|_{\Delta m_{\chi}} \propto \int\exp
\left(-\frac{\chi^2}{2}\right)\,{\rm d}^3{\bf P}_{\rm bkg},
\end{equation}
where $\chi^2$ is the chi-squared value of the model with background
parameters ${\bf P}_{\rm bkg}$. The lower limit of $m_{\chi}\tau$ 
at the 95\% credible interval is then obtained by setting
\begin{equation}
\frac{\int_0^{(m_{\chi}\tau)^{-1}}{\mathcal P}(x^{-1})\,dx}
{\int_0^{\infty}{\mathcal P}(x^{-1})\,dx}=0.95,
\end{equation}
in which $x=(m_{\chi}\tau)^{-1}$. The parameter space excluded by the 
EGB data is shown by the shaded region in Figure~\ref{fig:EGcons}. 
One can see that for $\chi\to\chi'ee$, $\chi\to\chi'\mu\mu$, and 
$\chi\to\chi'e\mu$ channels, there exists regions of parameter space 
where the AMS-02 positron excess can be explained without being excluded 
by the  EGB constraints. In particular, the $\chi\to\chi'ee$ 
channel is the most promising decay, after including the systematic 
uncertainties from the CR propagation models. For the cases with $\tau$ 
leptons and quarks, the EGB data strongly constrain the parameter space 
to explain the $e^+e^-$ excesses. It is worth mentioning that the non-smooth 
95\% limit is due to the two main peaks from the prompt and IC components 
and the weak structures of the data\footnote{The inclusion of the DM 
contribution, e.g., for $\chi'ee$, $\chi'\mu\mu$, and $\chi'e\mu$ channels, 
can actually give better fit to the data than the background. However, the 
decrease of the $\chi^2$ value is at most about 7.4, which corresponds to 
a $\sim2.2\sigma$ significance when adding two free parameters.}.

\section{The $\gamma$-ray fluxes from inner Galaxy}
\label{sec:GC}

In this section we discuss the compatibility of the DM models, that survived 
the EGB constraints, with the $\ga$-ray observation from the inner Galaxy.
An excess of $\ga$-rays between $1-10$ GeV from the inner Galaxy was 
identified in the Fermi-LAT data \cite{Goodenough:2009gk,Vitale:2009hr,
Hooper:2010mq,Hooper:2011ti,Abazajian:2012pn,Huang:2013pda,Gordon:2013vta,
Hooper:2013rwa,Abazajian:2014fta,Daylan:2014rsa,Zhou:2014lva,Calore:2014xka,
Huang:2015rlu,TheFermi-LAT:2015kwa}. The morphology and energy spectrum
of this excess has been shown to be consistent with ${\cal O}(10)$ GeV 
DM annihilation with a cross section consistent with thermal production. 
However, there are also astrophysical alternatives which can explain the 
data \cite{Abazajian:2010zy,Mirabal:2013rba,Yuan:2014rca,Petrovic:2014uda,
Petrovic:2014xra,Carlson:2014cwa,Cholis:2015dea,Carlson:2016iis,
Lee:2015fea,Bartels:2015aea}. Since the mass scale of the DM particle
relevant to the GC excess is different from that to explain the positron
excess, which is the focus of this work, we use the GC $\ga$-ray flux
as a consistency check on our model. 

The total DM induced $\gamma$-ray fluxes from the inner Galaxy include
both the prompt emission and the secondary IC and bremsstrahlung emission. 
The prompt emission dominates at high energies, while the secondary
emission is more important at low energies. For the prompt emission, we
use the same method as the case of anti-GC (Eq.~(\ref{anti-GC})) to 
calculate its flux. The secondary emission is more complicated. Since
the inner Galaxy region contains a large part of the diffusion halo of
CRs, the diffusion and cooling of electrons in the Galactic optical/infrared
radiation field is important, and the secondary emission can not be
simply calculated in the same way as that in the extragalactic space 
and the anti-GC region in Sec. 4~\cite{Zhang:2009kp}. We split the 
calculation of the secondary emission into two parts. The first part 
is the region inside the propagation cylinder, for which we use GALPROP 
to calculate the propagation of electrons/positrons and their secondary 
emission. The second part is the region outside the propagation cynlinder, 
for which we adopt the same method as the calculation of the anti-GC emission.

\begin{figure}[!htb!]
\centering
\includegraphics[width=0.45\textwidth]{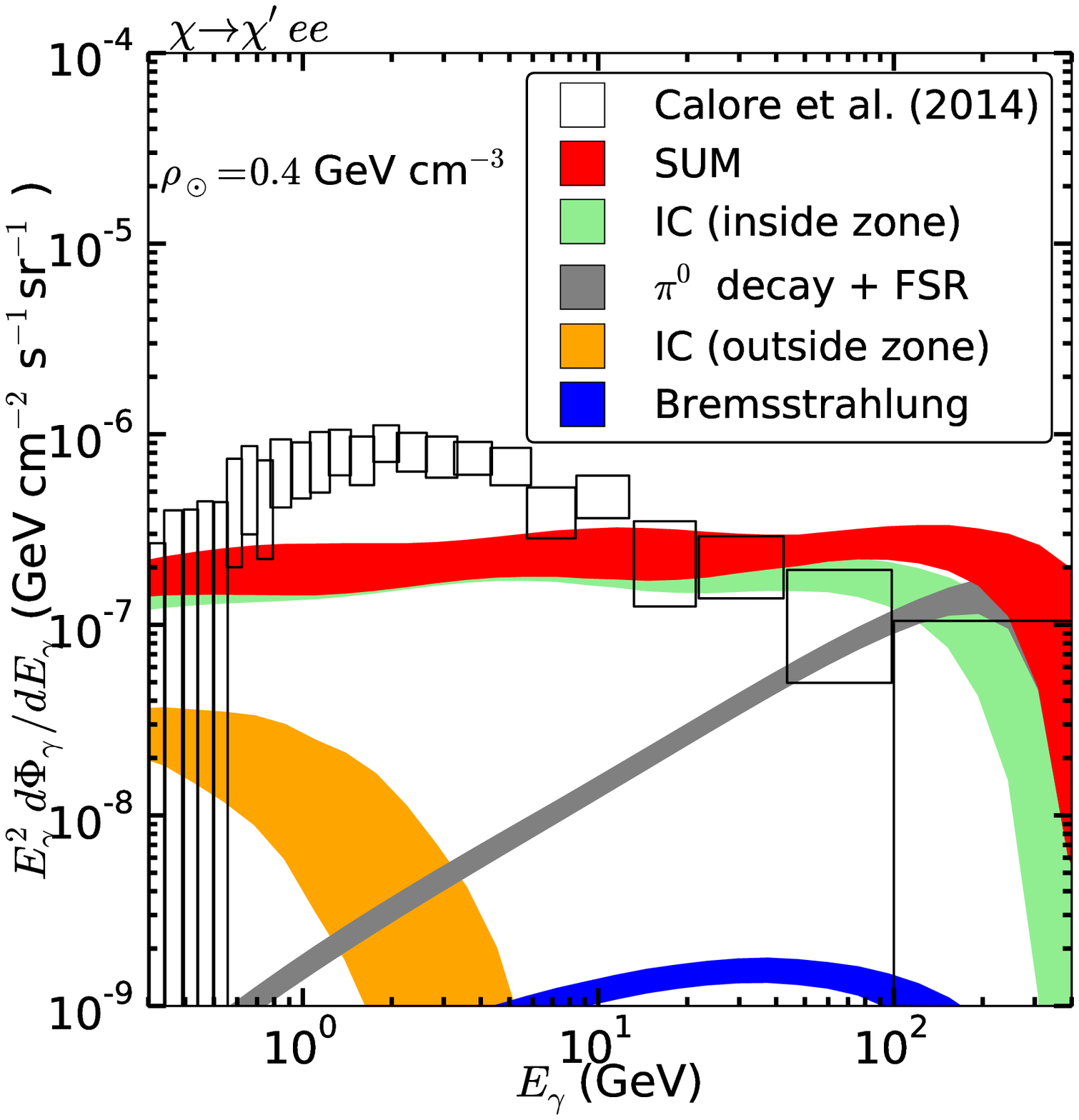}
\includegraphics[width=0.45\textwidth]{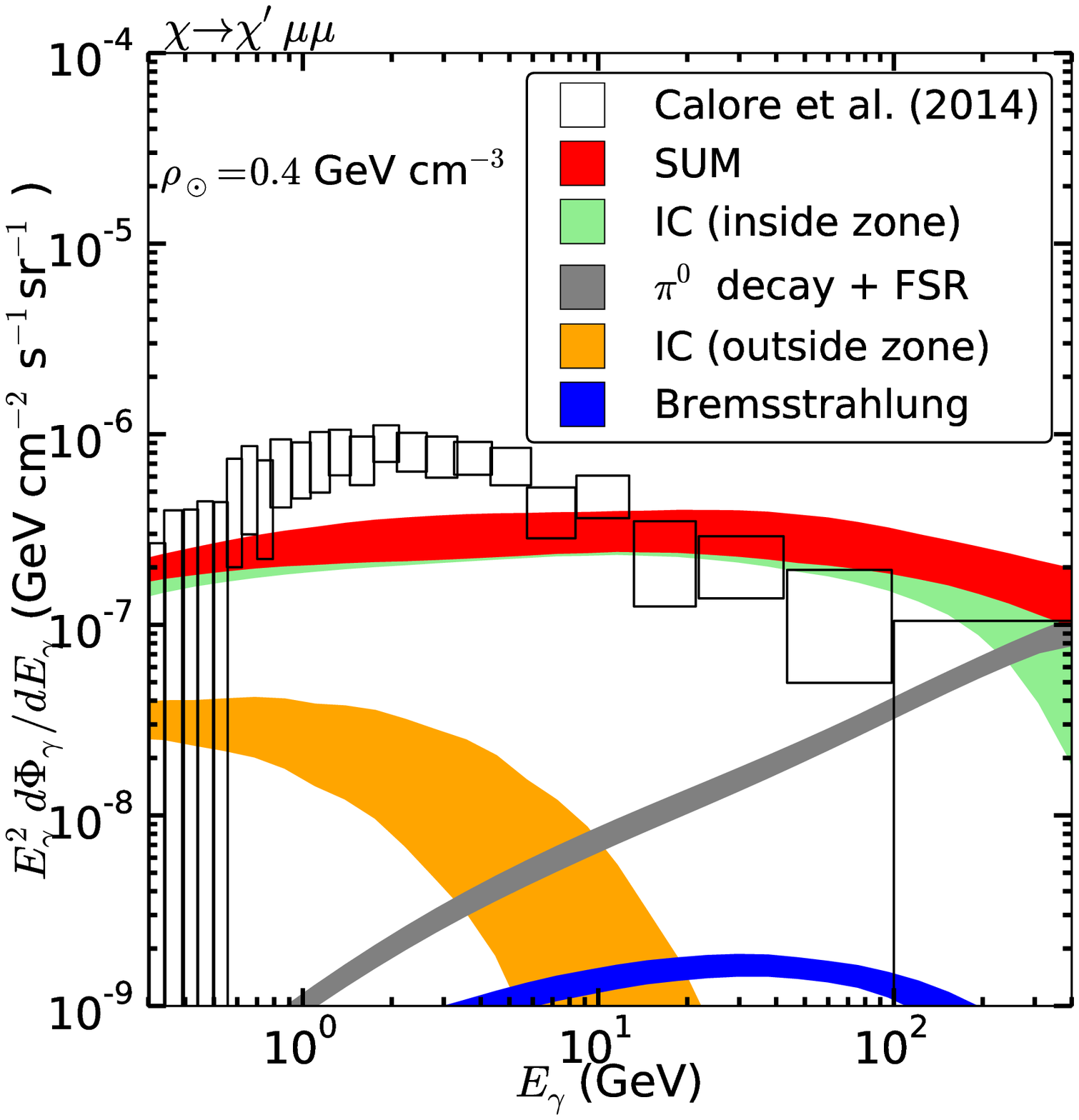}
\caption{The $\ga$-ray energy spectra based on the $95\%$ allowed 
parameters ($\Delta \mx$, $\tau\times \mx$) as presented in 
Figure~\ref{fig:EGcons}. The left panel is for $\chi\to\chi^\prime ee$, 
and the right panel is for $\chi\to\chi^\prime \mu\mu$. The two 
components of IC contributions from outside and inside the propagation
halo are presented by orange and light green bands, respectively. 
The bremsstrahlung is shown in blue, and the direct contribution from
the DM decay is shown in grey. The sum of the total contributions 
are coloured in red. The black boxes are the excess fluxes given in 
Ref.~\cite{Calore:2014xka}. 
}
\label{fig:AMS_GCE}
\end{figure}

Figure~\ref{fig:AMS_GCE} shows the computed $\ga$-ray fluxes based on the 
$95\%$ allowed ranges of parameters ($\Delta \mx$, $\mx\tau$) (within countours of Figure~\ref{fig:EGcons}) obtained
through fitting to the positron data, for $\chi\to\chi^\prime ee$
(left panel) and $\chi\to\chi^\prime \mu\mu$ (right panel) channels.
Compared with the data~\cite{Calore:2014xka} of the GC excess, we find
that, for the three-body decaying DM capable of explaining the AMS-02 
positron excess, the total $\ga$-ray fluxes are below those necessary to 
account for the inner galaxy excess except in the two highest energy bins. 
We should mention that there are still large systematic uncertainties on 
the fluxes of the GC excess due to different assumptions of the background
CR source distribution~\cite{TheFermi-LAT:2015kwa}, and there could be 
a high energy tail of the excess emission~\cite{Linden:2016rcf}. 
Therefore we conclude that the three-body decaying DM model (with decays to electrons or muons) is marginally 
consistent with the $\ga$-ray observations. It is worth mentioning that 
both the box-shaped total fluxes and the morphology, which could be altered 
due to the diffusion process~\cite{Lacroix:2015wfx}, from the DM decay are 
different from the data. Additional sources, such as millisecond 
pulsars~\cite{Abazajian:2010zy,Mirabal:2013rba,Yuan:2014rca,Petrovic:2014xra}, 
together with the decaying DM, might be responsible for the total GC excess.

\section{Conclusions}

In this work, we considered the possibility of using three-body decaying 
DM to explain the AMS-02 positron excess and examined the model compatibility 
with the $\ga$-ray/anti-proton constraints from Fermi-LAT/AMS-02, assuming 
the DM decays into a stable neutral particle and a pair of SM fermions.
The analysis is carried out using a scalar DM as an illustration, after 
demonstrating that  the secondary electron and photon spectra are insensitive 
to the assumptions on the specific quantum number of DM as well as the type of
interactions responsible for the DM decay.

We first investigated the region of parameter space which can account 
for the AMS-02 positron excess. To fit the positron data, leptonic 
channels require the mass splitting between decaying DM and the stable 
neutral partner to be of $\mathcal{O}(\text{TeV})$. For the hadronic 
channels, the mass splitting is pushed to be very large, primarily due 
to the constraints from the anti-proton data. Then we check both the 
Galactic and extragalactic $\ga$-ray data from Fermi-LAT observations. 
We find that channels which hadronize, e.g., $\chi\to\chi^\prime\tau\tau$ 
and $\chi\to\chi^\prime b\bar{b}$, will overshoot the EGB data observed
by Fermi-LAT. Primary decays into the electron and muon channels can 
survive, at least partly, from the EGB constraints. We finally check the 
$\ga$-ray emission of the DM decay into electron and muon channels in the 
inner Galaxy region, and showed that the predicted $\ga$-ray fluxes are 
below the observed excess in the inner Galaxy region considering the 
systematic uncertainties of the data. In the end, DM models with the 
decay channel of $\chi\to\chi^\prime ee$ could explain the AMS-02 
positron excess, without conflicting with the existing anti-proton and 
$\ga$-ray data.

Finally, we would like to point out that apart from continuous 
$\ga$-ray spectra, the three-body decaying DM can also radiatively 
produce monochromatic $\ga$-rays by connecting two final charged leptons 
of the same flavor into a loop with a photon insertion as studied in 
Ref.~\cite{Garny:2010eg}. Assuming no mass suppression due to the chirality 
flip from closing the fermion loop, the ratio of the partial decay width of 
$\chi \to \chi^\prime \, \ga$ to that of $\chi \to \chi^\prime \bar{f}f$ 
can be estimated as $\alpha/4 \pi \sim 10^{-3}$ with $\alpha$ being the 
fine structure constant. The best-fit values of the DM mass splitting 
responsible for the AMS-02 positron excess are between 0.6 and 1.1 TeV for 
the $e$ and $\mu$ final states, which correspond to the energies of 
monochromatic photons of 300 GeV to 1.1 TeV depending on $m_{\chi}$. 
The lower bounds on the DM lifetime of $\chi\to\ga\nu$ final state from 
the Fermi-LAT observations are about $10^{29}$ sec for DM masses of sub-TeV 
to TeV~\cite{Ackermann:2015lka}, which can be translated into $10^{26}$ sec 
for our case of $\chi\to\chi'\bar{f}f$, or even weaker if there is 
additional mass suppression for the radiative photon decay. On the other
hand, the best-fit value of the DM lifetime in our model is around $10^{27}$ 
sec, hence the current monochromatic photon bounds can be satisfied.
Future experiments such as the Cherenkov Telescope Array 
\cite{Bergstrom:2012vd} and the High Energy cosmic-Radiation Detection
facility \cite{Huang:2015fca} are expected to improve the line search
sensitivities effectively.
 
\section*{Acknowledgments}
H.-C. C. is supported in part by the US Department of Energy grant 
DE-SC-000999. I. L. is supported in part by the U.S. Department of 
Energy under Contract No. DE-AC02-06CH11357 at Argonne and Contract 
No. DE-SC0010143 at Northwestern. Y.S.T. was supported by World Premier 
International Research Center Initiative (WPI), MEXT, Japan.
W.C.H was supported by DGF Grant No. PA 803/10-1. Q.Y. is supported
by the 100 Talents program of Chinese Academy of Sciences.
H.-C. C. and I. L. would like to thank the hospitality of the Aspen 
Center for Physics, which is supported by the National Science Foundation 
under Grant No. PHYS-1066293, and the Kavli Institute for Theoretical 
Physics China at the Chinese Academy of Sciences, where part the work 
was completed. H.-C. C. would also like to acknowledge the hospitality 
of Academia Sinica in Taiwan.

%%%%%%%%%%%%%%%%%%%%%%%%%%%%%%%%%%%%%%%%%%%%%%%%%%%% 

\section*{Appendix A}

Marginalized posterior distributions of all the 12 model parameters 
from fitting to the charged CR data. Here $\chi\to\chi'ee$ channel and 
propagation model 2 are adopted for illustration. The diagonal panels 
show the 1-dimensional probability distributions, and other panels show 
the 2-dimensional $68\%$ (inner) and $95\%$ (outer) credible region.
The top (bottom) plot of the first column, for instance, corresponds 
to the 1-dimensional (2-dimensional) credible distribution for $\log A_e$ 
($\Phi_{\text{pb}}$ and $\log A_e$), with the rest parameters marginalized. 
Parameters from left to right are: the logarithm of the normalization of 
the locally observed electron flux at 25 GeV, the low rigidity spectral 
index, the logarithm of the first break rigidity, the medium rigidity 
spectral index, the logarithm of the second break rigidity, the high 
rigidity spectral index of the injection electron spectrum, the 
re-adjustment factor of secondary positrons, the re-adjustment factor of 
secondary anti-protons, the logarithm of $\Delta m_{\chi}$, the logarithm 
of $m_{\chi}\tau$, the solar modulation potential of electrons/positrons, 
and the solar modulation potential of anti-protons. 

\begin{figure}[!htb!]
\centering
\hspace*{-1.5cm}
\includegraphics[height=1.0\textwidth ,width=1.2\textwidth]{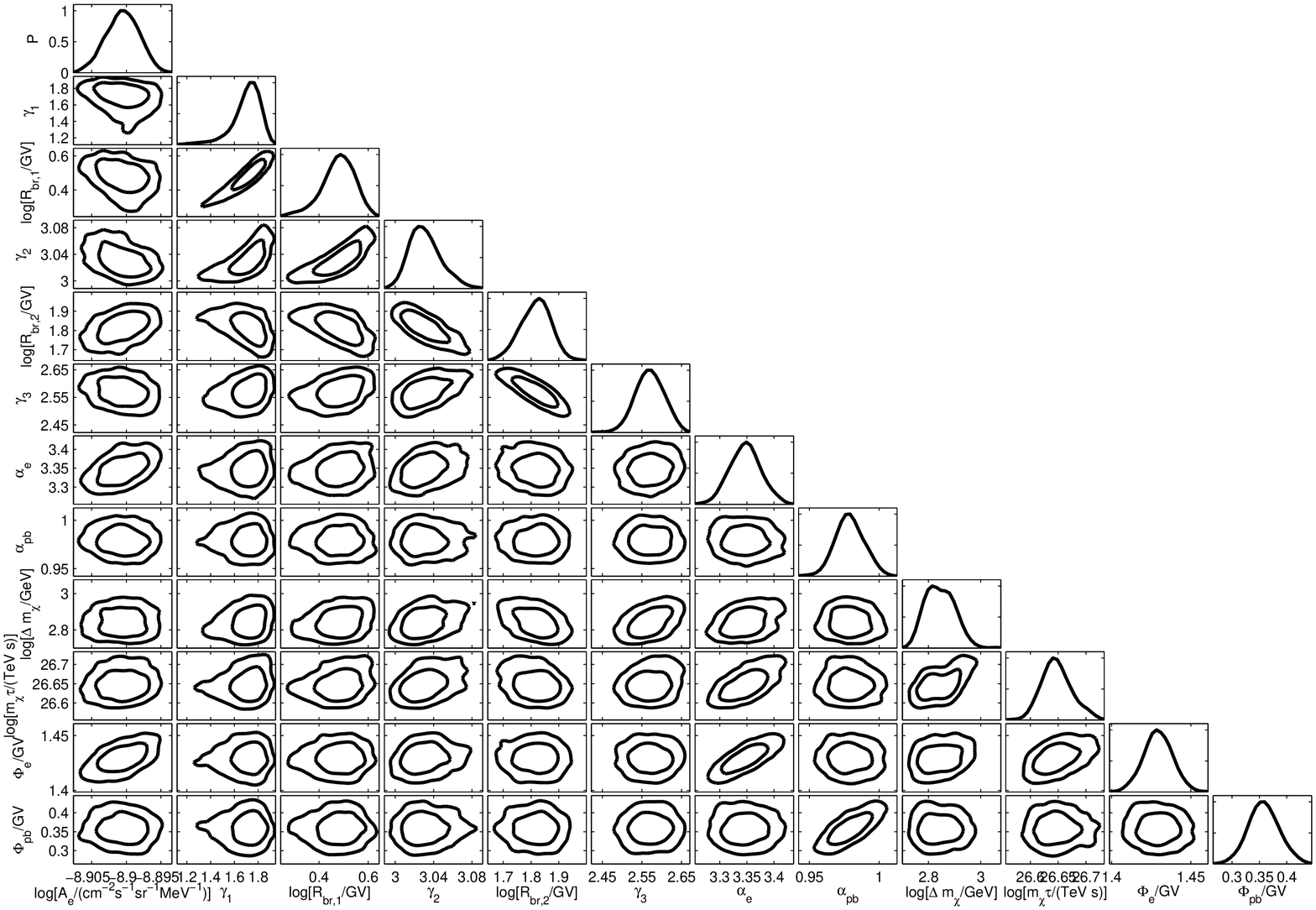}
\vspace*{-2.0cm}
\caption{Triangle plot to show the fitting distributions of the model
parameters for $\chi\to\chi'ee$ channel and propagation model 2.} 
\label{fig:tri_e}
\end{figure}

\section*{Appendix B}

\begin{figure}[!htb!]
\centering
\includegraphics[width=0.7\textwidth]{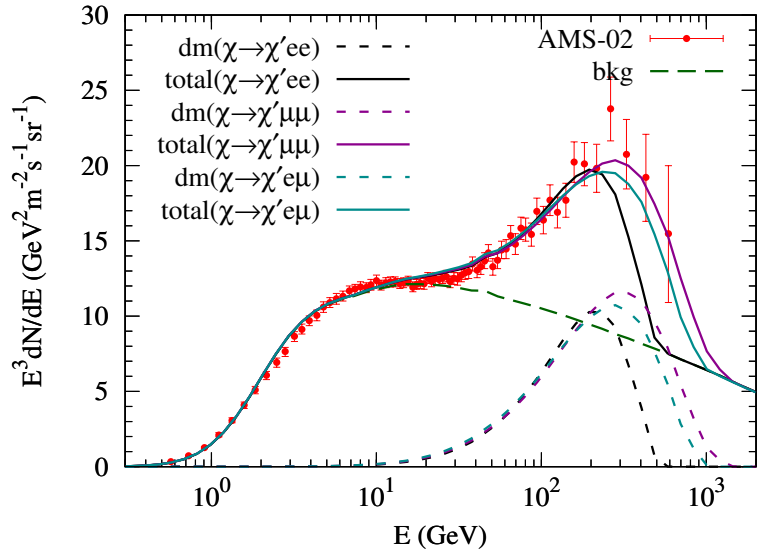}
\caption{The positron fluxes from our model fittings for $\chi\to\chi'ee$,
$\chi\to\chi'\mu\mu$, and $\chi\to\chi'e\mu$ channels, compared with the
five year data release of AMS-02 \cite{2016-AMS02-CERN}.}
\label{fig:posi_new}
\end{figure}

We have been aware of the five year data release of AMS-02 on December
8th, 2016, which reported updated measurements of the positron fluxes
to higher energies \cite{2016-AMS02-CERN}. We compare in 
Figure~\ref{fig:posi_new} the newest measurements with our model predictions
for $\chi\to\chi'ee$, $\chi\to\chi'\mu\mu$, and $\chi\to\chi'e\mu$ 
channels, which survive from the EGB constraints. We find that the 
latter two cases give good fittings to the data, while the extrapolation 
of the best-fit curve of the decay into an electron-positron pair does 
not well describe the new data points at higher energies. 

\bibliography{AMS_3b}
\bibliographystyle{JHEP}

\end{document}